\documentclass[12pt,english]{article}
\usepackage[LGR,T1]{fontenc}
\usepackage[latin9]{inputenc}
\usepackage{geometry}
\usepackage{array}
\usepackage{booktabs}
\usepackage{amsmath}
\usepackage{graphicx}
\usepackage{setspace}
\makeatletter

\DeclareRobustCommand{\greektext}{%
  \fontencoding{LGR}\selectfont\def\encodingdefault{LGR}}
\DeclareRobustCommand{\textgreek}[1]{\leavevmode{\greektext #1}}
\DeclareFontEncoding{LGR}{}{}
\DeclareTextSymbol{\~}{LGR}{126}
\newcommand{\lyxmathsym}[1]{\ifmmode\begingroup\def\b@ld{bold}
  \text{\ifx\math@version\b@ld\bfseries\fi#1}\endgroup\else#1\fi}

\providecommand{\tabularnewline}{\\}

\date{}
\usepackage{cite}
\usepackage{amsmath}

\makeatother

\usepackage{babel}
\begin{document}
\begin{center}
\textbf{\large{}Inclusion of enclosed hydration effects in the binding
free energy estimation of Dopamine D3 receptor complexes}
\par\end{center}{\large \par}

\begin{center}
Rajat Kumar Pal$^{1,2}$, Satishkumar Gadhiya$^{3,4}$, Pierpaolo
Cordone$^{2,4}$, Steven Ramsey$^{2,5}$, Lauren Wickstrom$^{6}$,
Wayne W. Harding$^{2,3,4}$, Tom Kurtzman$^{2,3,5}$ and Emilio Gallicchio$^{1,2,3,*}$
\par\end{center}

\textbf{1} Department of Chemistry, Brooklyn College, 2900 Bedford
Avenue, Brooklyn, NY 11210, USA

\textbf{2} PhD Program in Biochemistry, The Graduate Center of the
City University of New York, New York, NY 10016, USA

\textbf{3} PhD Program in Chemistry, The Graduate Center of the City
University of New York, New York, NY 10016, USA

\textbf{4} Department of Chemistry, Hunter College, 695 Park Avenue,
NY 10065, USA

\textbf{5} Department of Chemistry, Lehman College, 250 Bedford Park
Blvd. West, Bronx, NY 10468, USA

\textbf{6} Department of Science, Borough of Manhattan Community College,
199 Chambers Street, New York, NY 10007, USA

{*} egallicchio@brooklyn.cuny.edu

\section*{Abstract}

Confined hydration and conformational flexibility are some of the
challenges encountered for the rational design of selective antagonists
of G-protein coupled receptors. We present a set of C3-substituted
(-)-stepholidine derivatives as potent binders of the dopamine D3
receptor. The compounds are characterized biochemically, as well as
by computer modeling using a novel molecular dynamics-based alchemical
binding free energy approach which incorporates the effect of the
displacement of enclosed water molecules from the binding site. The
free energy of displacement of specific hydration sites is obtained
using the Hydration Site Analysis method with explicit solvation.
This work underscores the critical role of confined hydration and
conformational reorganization in the molecular recognition mechanism
of dopamine receptors and illustrates the potential of binding free
energy models to represent these key phenomena.

\section*{Introduction}

One critical aspect of molecular recognition is the change in the
hydration structure and hydration energetics induced by ligand binding.\cite{de2010role,Li2007,mancera2007molecular,ball2008water,ladbury1996just}
Water molecules trapped, for example, in hydrophobic pockets within
the binding site can be energetically disfavored as well as entropically
frustrated relative to bulk water. Hence, displacements of these water
molecules by the ligand can significantly enhance binding.\cite{nguyen2014thermodynamics,Setny2010,Haider2016hsa,Pal2017}
These effects are particularly important when comparing a series of
ligands of interest which differ in the way they displace enclosed
water molecules. The rational design of ligands using these principles
can lead to improvements of binding potency and receptor selectivity.\cite{Beuming2009} 

There have been significant efforts towards the development of methodologies
to model the thermodynamic parameters and structural properties of
water molecules at the protein surfaces.\cite{Young2007,huggins2012application,ross2012rapid,Ross2015,bodnarchuk2014strategies,sindhikara2013analysis}
Most of these methods employ an explicit representation of the solvent,
which is considered the ``gold standard'' for modeling macromolecular
complexes in part because of the capability of accurate representation
of specific hydration environments. It is challenging, however, to
access the time scales required to sample the changes in hydration
states and capturing the effects of water explusion from protein binding
sites induced by ligand binding.\cite{Graham2017,Ross2015,Michel2009a,Macdonald}
We have shown that the influence of confined hydration can be also
represented by a hybrid implicit solvent model trained on Hydration
Site Analysis (HSA)\cite{nguyen2014thermodynamics,Haider2016hsa}
data obtained with explicit solvation.\cite{Pal2017} The primary
purpose of this work is to explore the applicability of our hybrid
implicit solvent approach to protein-ligand systems. The dopamine
D3 receptor is an important medicinal target in which the ligand recognition
mechanism is heavily influenced by hydration effects. Due to conformational
variability, the complexities of hydration and molecular interaction
networks, and the lack of extensive structural information, it has
been very challenging, using conventional drug design and modeling
approaches, to design selective antagonists against the dopamine D3
family of receptors. We believe that molecular dynamics free energy
approaches combined with accurate modeling of hydration can be helpful
in the design of more effective and more specific antagonists.\cite{heidbreder2010current,Gadhiya2018,madapa2016synthesis,GPCRwaterPathway2014}

Dopamine D3 receptors, which are part of the G-protein coupled receptor
superfamily, are increasingly important as drug targets for the treatment
of a number of pathological conditions such as Parkinson's disease,
schizophrenia and drug abuse.\cite{maramai2016dopamine,Volkow2007,Brooks2000}
Dopamine receptors are classified under two families and five sub-types:
the D1 family, comprising the D1R and D5R receptors which stimulate
the production of cAMP, and the D2 family, comprising the D2R, D3R
and D4R receptors which have inhibitory functions in cAMP production
and downstream signaling. While both these receptor families have
been targeted for the treatment of neurological disorders, it has
been challenging to design specific antagonists within the D2 receptor
subfamily. Most of the drugs tested act as dual D2/D3 antagonists.\cite{Chien2010,cho2010current,InsightsD1Ragonist2012li,stepholidine2007clinicalMo}
D2 receptor antagonism has been associated with serious neurological
side effects.\cite{extrapyramidal1997knable,extrapyramidal2017D2skyes}
D3 receptors, on the other hand, which also have high affinity towards
dopamine were observed to significantly affect synaptic transmission
and can be potential targets in the treatment of neurological disorders
, especially related to drug addiction and craving responses.\cite{song2014blockade,cho2010current,medicationTargets2010D3}

The mechanism of antagonism of D3 receptors has been intensely studied
to gain an understanding of how to develop potent and selective antagonists.\cite{Chien2010,D3crypticpocket2017ferruz,Gadhiya2018,molecularDeterminantD32012newman,InsightsD1Ragonist2012li}
The crystal structure of the D3 receptor in complex with eticlopride,\cite{Chien2010}
a dual D2/D3 antagonist, has been very helpful in understanding the
intermolecular interactions in the orthosteric binding site (OBS)
of the D3 receptor. It also revealed a secondary binding site (SBS)
which is believed to be a critical molecular recognition site. A recent
study has also suggested the existence of a cryptic pocket in the
orthosteric binding site (OBS) of the dopamine D3 receptor.\cite{D3crypticpocket2017ferruz}
These important discoveries have provided valuable information for
the development of D3 selective ligands.\cite{Gadhiya2018,madapa2016synthesis}

\begin{figure}
\begin{centering}
\includegraphics[scale=0.2]{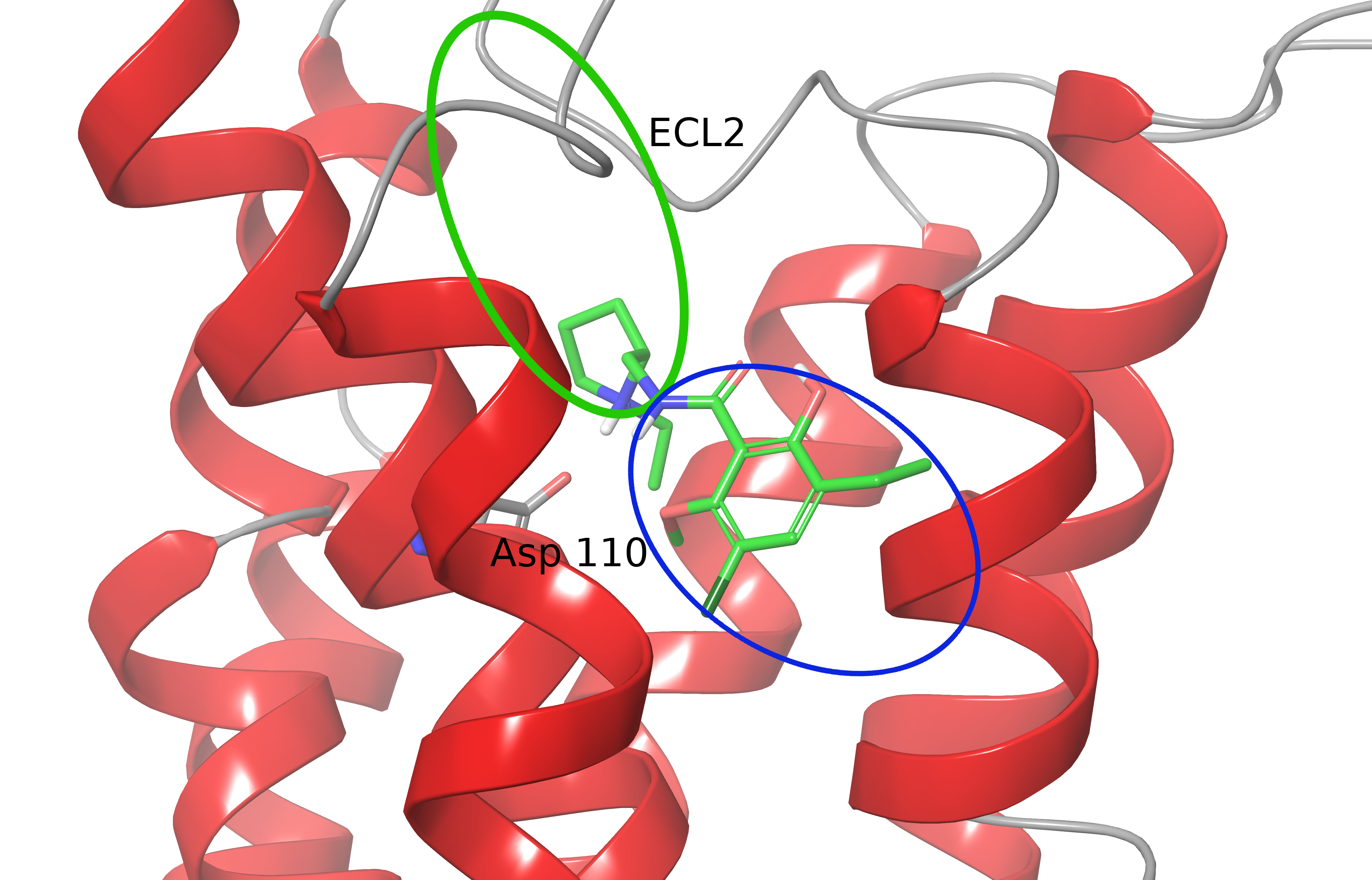}
\par\end{centering}

\caption{Crystal structure of the dopamine D3 receptor with Eticlopride bound
at the binding site \cite{Chien2010}. This representation shows the
approximate position of the orthosteric binding site (OBS) with a
blue oval and the secondary binding site (SBS) with a green oval.\label{fig:Crystal-structure-of-D3}}
\end{figure}

The orthosteric binding site (OBS) of D3 is surrounded by the helices
III, V, VI and VII comprising Ser 192$^{5.42}$, Ser 193$^{5.43}$,
Ser 196$^{5.46}$, Cys 114$^{3.36}$, His 349$^{6.55}$, Phe 345$^{6.51}$,
Phe 346$^{6.52}$ and Val 189$^{5.39}$ residues. The secondary binding
site (SBS), also referred as the extracellular extension, is located
at the interface of helices I, II, III, VII and the extracellular
loops ECL1 and ECL2 (Fig.~\ref{fig:Crystal-structure-of-D3}). The
OBS is conserved in both D2 and D3 receptors but differ in the residue
composition at the SBS. As exemplified by the structure of D3 bound
to eticlopride\cite{Chien2010} (Fig.~\ref{fig:Crystal-structure-of-D3}),
the interaction of ligands to the OBS of D3 is characterized by a
salt-bridge between the carboxylate group of Asp 110$^{3.32}$ in
helix III of D3 and the protonated amine group of eticlopride. This
salt-bridge interaction is believed to be pharmacologically crucial
in binding of ligands at the OBS of dopamine D3 receptor and to other
dopaminergic receptors.\cite{Chien2010} Previous studies have highlighted
the challenges of designing specific antagonists against the dopamine
D3 receptor.\cite{molecularDeterminantD32012newman,madapa2016synthesis,heidbreder2010current}

\begin{figure}
\centering{}\includegraphics[scale=0.5]{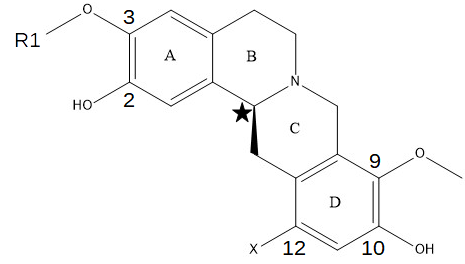}\caption{Structure of the (-)-stepholidine core with four rings annotated alphabetically
as referenced in the text. R1 represents the substitution at the C3
position. The chiral carbon is labeled by a star.\label{fig:Stepholidine_structure}}
\end{figure}

In this study, we focus on the interaction of the D3 receptor with
a series of derivatives of (-)-stepholidine (Table~\ref{tab:Stepholidine_derivatives_schematic}),
a natural product displaying dual D1 and D2 activity and observed
to have antipsychotic activities.\cite{stepholidine2007clinicalMo,stepholidine2015panDopamine,stepholidine2017mtorBzhang,stepholidine2018manuszak}
Motivated by the previous work on the synthesis and activity of the
(-)-stepholidine C9 derivatives\cite{madapa2016synthesis} aimed at
achieving a dual D1/D3 activity, we continued our Structure-Activity
Relationship (SAR) studies using the tertrahydroprotoberberine (THPBs)
scaffold to synthesize a new set of compounds targeting the dopamine
receptors. In comparison to the compounds previously assayed which
are substituted with alkyl chains at the C9 position of the THPB scaffold,
compounds synthesized and studied in this work are substituted at
the C3 position (Fig.~\ref{fig:Stepholidine_structure} and Table~\ref{tab:Stepholidine_derivatives_schematic}).
The motivation of synthesis and substitution at the C3 position is
to extend these molecules to access the secondary binding site (SBS)
which have the potential to improve receptor selectivity for these
compounds.\cite{madapa2016synthesis} Due to the lack of a crystal
structure, the mode of interaction of (-)-stepholidine derivatives
with the D3 receptor remains uncertain.\cite{madapa2016synthesis,D1D2agonist2007Fu,InsightsD1Ragonist2012li}

In this work, we report the first assessment of a novel computational
strategy by using an implicit solvent model to model the effects of
water expulsion in protein-ligand binding. This is done by acquiring
the thermodynamic properties of binding site water molecules in dopamine
D3 receptor from explicit solvent simulations and estimating the binding
free energies of the complexes of (-)-stepholidine analogues' with
the D3 receptor by incorporating hydration parameters in an implicit
solvent model. This allowed us to capture localized enclosed hydration
effects which could not be captured by using conventional descriptions
of solvation.

\begin{table}
\caption{List of the (-)-stepholidine derivatives considered in this work.
All substitution are made at the C3 position of the (-)-stepholidine
core as shown in Fig.~\ref{fig:Stepholidine_structure}. \label{tab:Stepholidine_derivatives_schematic}}

\centering{}%
\begin{tabular*}{0.8\columnwidth}{@{\extracolsep{\fill}}ccc}
\hline 
\multicolumn{3}{c}{(-)-stepholidine C3 derivatives}\tabularnewline
\hline 
 & {\footnotesize{}x} & {\footnotesize{}R1}\tabularnewline
\hline 
{\footnotesize{}1a} & {\footnotesize{}H} & {\footnotesize{}Et}\tabularnewline
{\footnotesize{}1b} & {\footnotesize{}H} & {\footnotesize{}n-Pr}\tabularnewline
{\footnotesize{}1c} & {\footnotesize{}H} & {\footnotesize{}n-Bu}\tabularnewline
{\footnotesize{}1d} & {\footnotesize{}H} & {\footnotesize{}n-Pen}\tabularnewline
{\footnotesize{}1e} & {\footnotesize{}H} & {\footnotesize{}n-Hex}\tabularnewline
{\footnotesize{}1f} & {\footnotesize{}H} & {\footnotesize{}2-fluoro ethyl}\tabularnewline
\hline 
\end{tabular*}
\end{table}

\section*{Methods}

\subsection*{Hydration Site Analysis of the binding site of the D3 receptor}

\begin{figure}
\begin{centering}
\includegraphics[scale=0.2]{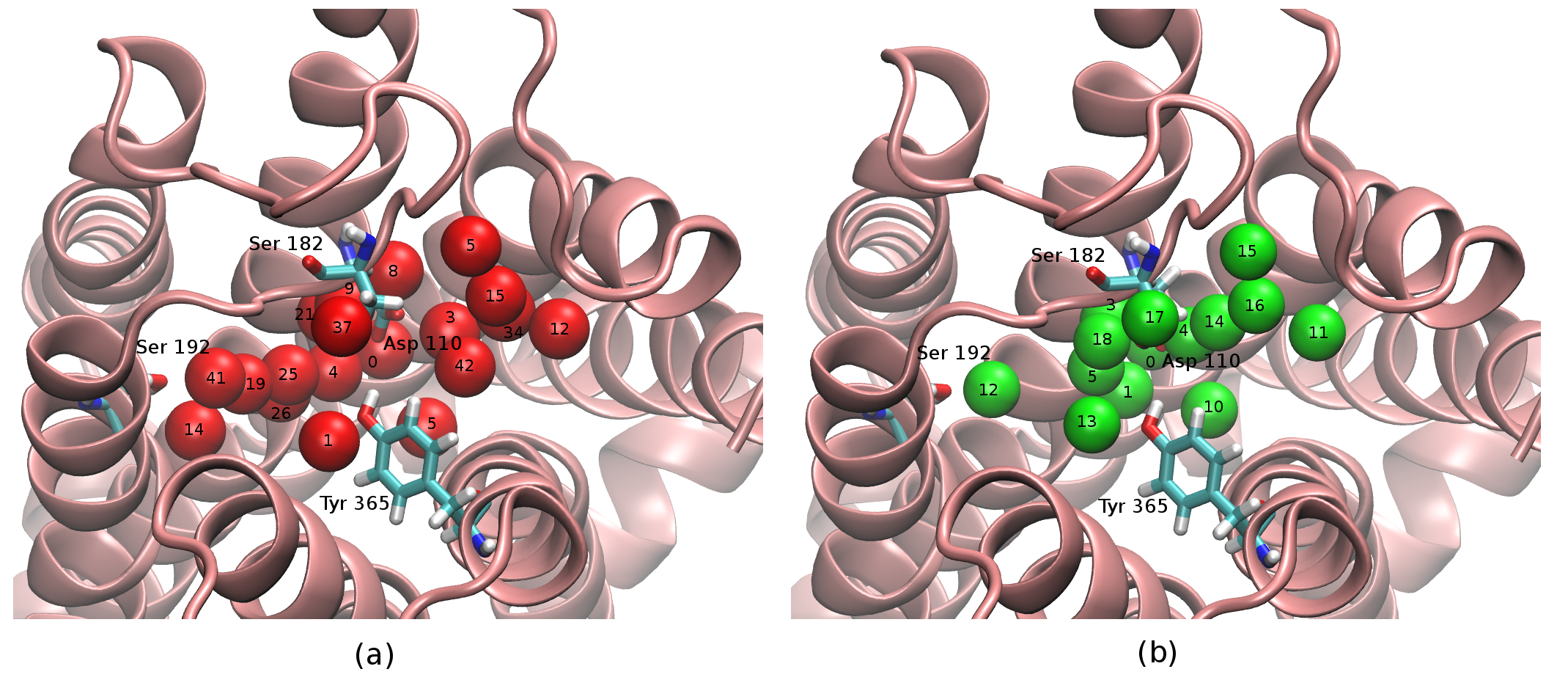}
\par\end{centering}

\caption{(a) Location of hydration sites (red) within the binding cavity of
the Dopamine D3 receptor as mapped by Hydration Site Analysis. (b)
Hydration spheres (green) of the AGBNP2 model for the same receptor
structure in (a). The positions of the AGBNP2 hydration spheres are
functions of the internal coordinates of the receptor. \label{fig:HSA_and_AGBNP2_sites}}
\end{figure}

The thermodynamic and structural properties of water molecules in
the binding site of the receptor were studied using the Hydration
Site Analysis (HSA) method.\cite{Young2007,Haider2016hsa} Briefly,
HSA is based on the analysis of molecular dynamics trajectories with
explicit solvation, whereby molecular dynamics simulations are performed
to identify regions with significant water density near the receptor
surface. Average thermodynamic quantities such as enthalpy, entropy
and free energies are calculated for these sites using the concept
of Inhomogeneous Solvation Theory.\cite{nguyen2014thermodynamics,Lazaridis:98}
HSA explicit solvent simulations are performed on a restrained receptor
structure. The trajectories are then processed to cluster hydration
site locations and analyzed for their thermodynamic estimates as described
elsewhere.\cite{nguyen2014thermodynamics,Haider2016hsa} The total
energy, $E_{{\rm total}}$ for each of these sites are calculated
as the sum of the one-half of the mean solute-water $E_{{\rm sw}}$
interaction energy and one-half of the mean water-water $E_{{\rm ww}}$
interaction energy. The excess energies of the hydration sites relative
to bulk value are used to classify them as either favorable or unfavorable
water sites. Unfavorable sites are those that, when displaced by the
ligand, are believed to enhance the binding affinity. The locations
and average solvation energies for each of the sites identified for
the D3 receptor are shown in Fig.~\ref{fig:HSA_and_AGBNP2_sites}a
and Table~\ref{tab:HSA-score}. 

Proteins can be highly dynamic. Hence, a single structure is often
an insufficient representation of the structural variability of the
hydration layer of a protein receptor. This is particularly so in
the present work, where the ligands we considered could induce different
conformations of the receptor when bound. To address conformational
variability, in this work, we obtained HSA hydration maps for a series
of D3 receptor structures obtained from induced-fit docking calculations
with different ligand types including the available crystal structure\cite{Chien2010}
(see Computational Details). The location and energies of the hydration
sites were averaged from all receptor conformations to obtain a single
hydration map as shown in Fig.~\ref{fig:HSA_and_AGBNP2_sites}a. 

The solvation energies and locations of the explicit hydration sites
were then used to position the hydration spheres of the AGBNP2 implicit
solvation model and to set their strengths (\ref{eq:HSA_score}).
The strength of the hydration spheres were set according to the HSA
scores

\begin{equation}
[E_{{\rm total}}(i)-E_{{\rm bulk}}]p(i)\label{eq:HSA_score}
\end{equation}
where $i$ is the index of the HSA hydration sites, $p(i)$ is the
water occupancy of the site, $E_{{\rm total}}(i)$ is the total energy
of the site and $E_{{\rm bulk}}$ is the corresponding reference value
obtained from OPC\cite{Izadi2014} neat water ($E_{{\rm bulk}}=-12.24$
kcal/mol). 

\subsection*{Parameterization of the AGBNP2 Enclosed Hydration Model}

Even slight variations in atomic positions are known to cause significant
changes in hydration structure.\cite{nguyen2012grid,Beuming2009,Oroguchi2016}
We attempted to capture specific ligand-induced conformational changes,
as well as thermal fluctuations of the hydration structure by considering
multiple structures of the D3 receptor (see Computational Details).
Hydration site maps were obtained individually for each of the three
receptor structures using HSA.\cite{Haider2016hsa} These hydration
maps were then integrated into a single hydration map (see Fig.~\ref{fig:HSA_and_AGBNP2_sites}a)
by averaging the free energy weights of neighboring hydration sites
from the individual maps. The energies and water occupancies of the
HSA hydration regions were used to obtain the enclosed hydration corrections
for the AGBNP2 hydration spheres using eq.~(\ref{eq:HSA_score}). 

\begin{table}[h]
\begin{centering}
\caption{Summary of the placement and parameterization of the AGBNP2 enclosed
hydration sites for the dopamine D3 receptor binding site. \label{tab:HSA-score}}

\par\end{centering}

\noindent \centering{}%
\begin{tabular}{>{\centering}p{0.1\textwidth}>{\centering}p{0.1\textwidth}>{\centering}p{0.1\textwidth}>{\centering}p{0.142\textwidth}>{\centering}p{0.04\textwidth}>{\centering}p{0.15\textwidth}c}
\hline 
{\footnotesize{}Location$^{a}$} & {\footnotesize{}HSA site Id$^{b}$} & {\footnotesize{}AGBNP2 site Id$^{b}$} & {\footnotesize{}AGBNP2 anchoring type $^{c}$} & {\footnotesize{}$p_{s}^{d}$} & {\footnotesize{}$(E-E_{{\rm bulk}})^{e}$} & {\footnotesize{}$(E-E_{{\rm bulk}}){\rm \times}p_{s}$$^{f}$}\tabularnewline
\hline 
{\footnotesize{}OBS} & {\footnotesize{}0} & {\footnotesize{}0,1} & {\footnotesize{}Asp 110 backbone carbonyl} & {\footnotesize{}$1.00$} & {\footnotesize{}$2.36$} & {\footnotesize{}$2.36$}\tabularnewline
{\footnotesize{}OBS} & {\footnotesize{}3,4,8,21} & {\footnotesize{}3,4,5} & {\footnotesize{}Asp 110 side chain carboxylate} & {\footnotesize{}$0.86$} & {\footnotesize{}$3.28$} & {\footnotesize{}$2.83$}\tabularnewline
{\footnotesize{}OBS} & {\footnotesize{}25,26} & {\footnotesize{}9} & {\footnotesize{}Center of mass} & {\footnotesize{}$0.66$} & {\footnotesize{}$1.92$} & {\footnotesize{}$1.27$}\tabularnewline
{\footnotesize{}OBS} & {\footnotesize{}14,19,41} & {\footnotesize{}12} & {\footnotesize{}Center of mass} & {\footnotesize{}$0.57$} & {\footnotesize{}$5.32$} & {\footnotesize{}$3.05$}\tabularnewline
{\footnotesize{}OBS} & {\footnotesize{}1} & {\footnotesize{}13} & {\footnotesize{}Center of mass} & {\footnotesize{}$0.87$} & {\footnotesize{}$2.80$} & {\footnotesize{}$2.44$}\tabularnewline
{\footnotesize{}OBS/SBS boundary} & {\footnotesize{}11} & {\footnotesize{}10} & {\footnotesize{}Center of mass} & {\footnotesize{}$0.83$} & {\footnotesize{}$2.32$} & {\footnotesize{}$1.92$}\tabularnewline
{\footnotesize{}OBS/SBS boundary} & {\footnotesize{}9} & {\footnotesize{}18} & {\footnotesize{}Ser 182 hydroxyl hydrogen} & {\footnotesize{}$0.87$} & {\footnotesize{}$0.21$} & {\footnotesize{}$0.19$}\tabularnewline
{\footnotesize{}SBS} & {\footnotesize{}12} & {\footnotesize{}11} & {\footnotesize{}Center of mass} & {\footnotesize{}$0.63$} & {\footnotesize{}$0.92$} & {\footnotesize{}$0.58$}\tabularnewline
{\footnotesize{}SBS} & {\footnotesize{}34,42} & {\footnotesize{}14} & {\footnotesize{}Center of mass} & {\footnotesize{}$0.58$} & {\footnotesize{}$1.31$} & {\footnotesize{}$0.77$}\tabularnewline
{\footnotesize{}SBS} & {\footnotesize{}5} & {\footnotesize{}15} & {\footnotesize{}Center of mass} & {\footnotesize{}$0.95$} & {\footnotesize{}$0.34$} & {\footnotesize{}$0.33$}\tabularnewline
{\footnotesize{}SBS} & {\footnotesize{}15} & {\footnotesize{}16} & {\footnotesize{}Center of mass} & {\footnotesize{}$0.68$} & {\footnotesize{}$0.50$} & {\footnotesize{}$0.34$}\tabularnewline
{\footnotesize{}SBS} & {\footnotesize{}37} & {\footnotesize{}17} & {\footnotesize{}Center of mass} & {\footnotesize{}$0.34$} & {\footnotesize{}$0.08$} & {\footnotesize{}$0.03$}\tabularnewline
\hline 
\end{tabular}\\
\begin{minipage}[t]{1\columnwidth}%
{\scriptsize{}$^{a}$OBS: Orthosteric binding site; SBS: Secondary
binding site. $^{b}$Site Id as shown in Fig.~\ref{fig:HSA_and_AGBNP2_sites}
$^{c}$See reference. $^{d}$Average water occupancy of the site measured
by HSA $^{e}$Average energy of the site relative to bulk measured
by HSA, $E_{{\rm bulk}}=-12.24$ kcal/mol. $^{f}$Overall energy score
of the HSA sites indicated in column 2 and of the enclosed hydration
score of the AGBNP2 hydration spheres indicated in column 3 in kcal/mol.}%
\end{minipage}
\end{table}

\begin{figure}
\begin{centering}
\includegraphics[scale=0.2]{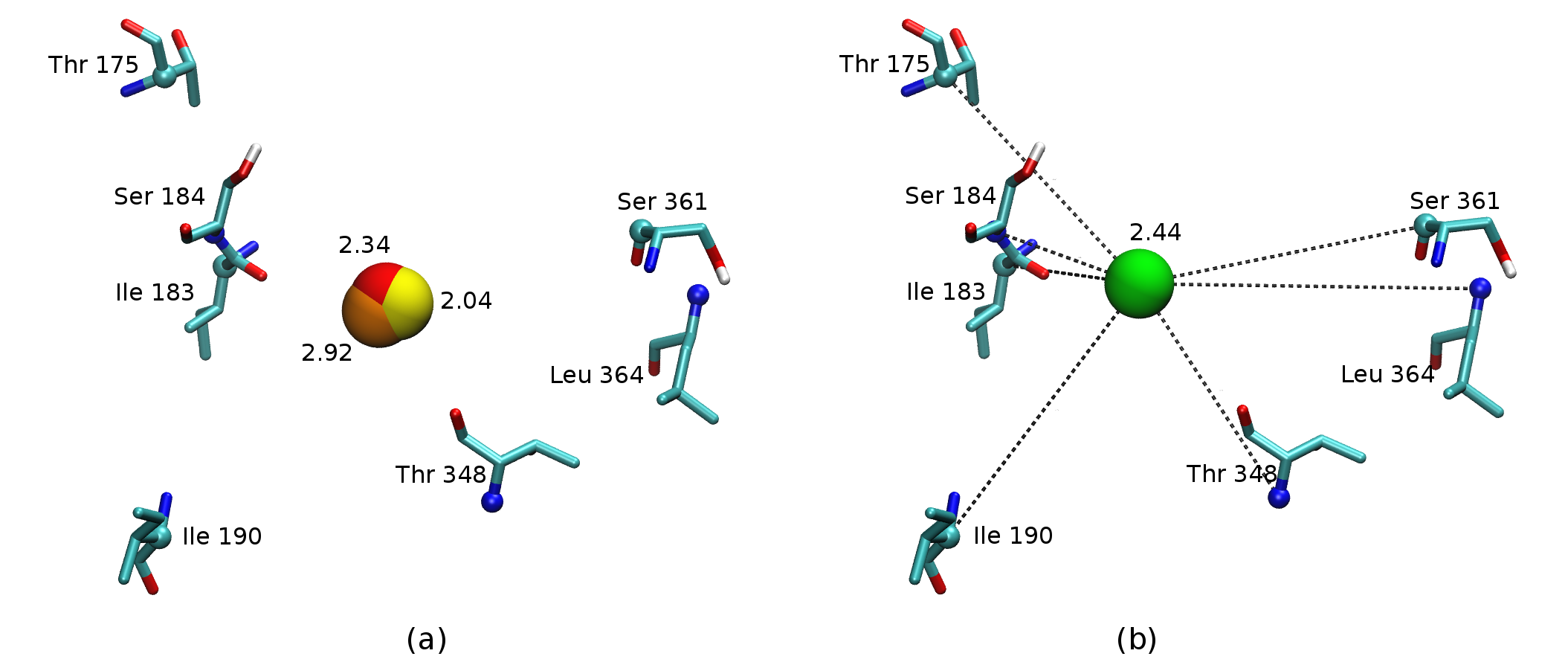}
\par\end{centering}

\caption{(a) Location of a hydration site identified by HSA using three receptor
structures (residues from one receptor structure shown for clarity);
the overlapping red, yellow and orange spheres represent a hydration
site identified by each receptor structure; the energetic penalties
incurred from each HSA map are annotated in kcal/mol, (b) An AGBNP2
hydration sphere (green) is placed and scored by averaging the energetic
penalties from the three maps at the location of the HSA site; the
AGBNP2 hydration sphere is placed at the geometrical center of the
atoms represented in CPK and is anchored to respective atoms during
the simulation.\label{fig:agbnp2-cm-site-example}}
\end{figure}

The energetically unfavorable hydration sites identified by HSA, and
thus good candidates for displacement by the ligand, were found to
be distributed throughout the dopamine D3 receptor binding site. These
were reproduced as best as possible with AGBNP2 hydration spheres
within the limitations of the available anchoring methods.\cite{Gallicchio2009,Pal2017}
To ensure translational and rotational invariance of the AGBNP2 implicit
solvation function, hydration spheres are located only in terms of
molecular internal coordinates, that is by specifying distance and
angle geometries in relation to selected atoms of the receptor. The
geometries that were employed most often in this work have been for
sites attached to polar hydrogen atoms and for sites anchored to mimic
the lone pair orbitals of carbonyl and carboxylate groups. When a
suitable anchoring geometry could not be found, AGBNP2 hydration spheres
have been positioned at the geometrical center of a group of atoms
of the receptor, typically backbone ${\rm C_{{\rm \alpha}}}$, ${\rm C_{{\rm {\rm \beta}}}}$
and ${\rm N}$ atoms (Fig.~\ref{fig:agbnp2-cm-site-example}).\cite{Pal2017}
The resulting AGBNP2 hydration sites are shown in Fig.~\ref{fig:HSA_and_AGBNP2_sites}b
and their parameterization are listed in Table~\ref{tab:HSA-score}. 

Because of the complexities of enclosed hydration phenomena and their
variations due to the motion of receptor atoms, it has been challenging
to formulate an unsupervised and automated protocol to map HSA results
to AGBNP2 spheres. Within the general framework outlined above, some
manual adjustments were made. One adjustment was made to model strongly
unfavorable HSA hydration sites (HSA site Ids - 3,4,8 and 21) identified
at hydrogen-bonding distance to the carboxylate group of the critical
Asp110$^{3.32}$ residue. Because AGBNP2 attaches eight equinergetic
solvation spheres to carboxylate groups,\cite{Gallicchio2009} we
decided to distribute the HSA excess energy of this site among the
three out of eight carboxylate hydration spheres of Asp110$^{3.32}$
with non-zero water occupancy. Adjustments were also made to treat
HSA hydration sites in close proximity to each other. Due to the limitations
in mapping accurately the position of AGBNP2 spheres, in these case,
we modeled nearby groups of HSA sites with a single AGBNP2 hydration
sphere by assigning to it the sum of the energy weights of each HSA
site as shown in Table~\ref{tab:HSA-score}.

\subsection*{Binding free energy model}

The protein-ligand complexes are modeled using the OPLS-AA/AGBNP2
effective potential, in which the OPLS-AA\cite{Kaminski:2001,Jacobson:Kaminski:Friesner:Rapp:2002}
force field defines the covalent and non-bonded inter-atomic interactions.
Solvation effects are modeled implicitly using the Analytic Generalized
Born plus non-polar (AGBNP2) model.\cite{Gallicchio2009} According
to this model, the hydration free energy $\Delta G_{{\rm h}}$ of
the receptor-ligand complex is computed as the sum of electrostatic
$\Delta G_{{\rm elec}}$, non-polar, $\Delta G_{{\rm np}}$, and short-range
solute-water interactions, $\Delta G_{{\rm hs}}$ : 

\begin{equation}
\Delta G_{{\rm h}}=\Delta G_{{\rm elec}}+\Delta G_{{\rm np}}+\Delta G_{{\rm hs}}\label{eq:hydration_free_energy}
\end{equation}

The electrostatic component of the hydration free energy is computed
using a modified continuum dielectric Generalized Born model.\cite{Qiu:Shenkin:Hollinger:Still:97,Hawkins:Cramer:Truhlar:96}
The non-polar component includes a surface-area dependent term that
accounts for the free energy of creating the solute cavity within
the solvent, and a Born-radius dependent term that accounts for long
range solute-solvent van der Waals interactions.\cite{Gallicchio2009}
In AGBNP2, short-ranged solute-solvent interactions, such as hydrogen
bonding are modeled by means of hydration spheres placed on the solute
surface. A geometrical procedure measures the water occupancy of each
hydration sphere, which is then used to weigh its contribution to
the solute hydration free energy according to the expression:

\begin{equation}
\Delta G_{{\rm hs}}=\sum_{s}h_{s}S(w_{s})\label{eq:h_bonding_component}
\end{equation}
where $w_{s}$ is the water occupancy factor of the sphere defined
as 

\begin{equation}
w_{{\rm s}}=\dfrac{V_{{\rm s}}^{{\rm free}}}{V_{{\rm s}}}\label{eq:occupancy_factor}
\end{equation}
where $V_{{\rm s}}$ is the volume of each sphere and $V_{{\rm s}}^{{\rm free}}$
is the volume of the portion of the sphere occupied by water. $S$
is a switching function that smoothly turns off an hydration sphere
if its water occupancy is below a given threshold. The $h_{s}$ parameter
measures the hydration strength of the corresponding hydration site.
Negative $h_{s}$ values describe hydration sites contributing favorably
to the hydration free energy, whereas positive values are used for
sites which contribute unfavorably to the hydration free energy.\cite{Pal2017}

In this study, almost all hydration sites identified by HSA inside
the binding site are energetically unfavorable. The strength of AGBNP2
hydration site spheres, thus having positive $h_{s}$ values are used
to define unfavorable water molecules in the binding site of the receptor,
which, when displaced by the ligand, contribute favorably to binding.
The $h_{{\rm s}}$ energy values are obtained from the explicit solvent
HSA analysis as described above and are listed in Table ~\ref{tab:HSA-score}. 

Absolute binding free energies of the dopamine D3 receptor bound to
(-)-stepholidine C3 analogues were calculated by means of a Single
Decoupling (SDM) binding free energy approach\cite{Gallicchio2010}
employing an alchemical potential energy function of the form:

\begin{equation}
U_{\lambda}(\mathbf{r})=U_{0}(\mathbf{r})+\lambda u(\mathbf{r})\label{eq:U_r-1}
\end{equation}
where $\mathbf{r}=(\mathbf{r}_{{\rm R}},\mathbf{r}_{{\rm L}})$ are
the atomic coordinates of the receptor-ligand complex, $U_{0}$ represents
the effective potential energy of the uncoupled complex when receptor
and ligand are not interacting (such as if they were at infinite separation),
$\lambda$ is the alchemical progress parameter which linearly couples
receptor and ligand through the binding energy function $u(\mathbf{r})$,
defined as the change in the effective potential energy of the complex
for bringing the receptor and ligand from infinite separation to the
conformation $\mathbf{r}$. Based on eq.~(\ref{eq:U_r-1}), the complex
is uncoupled at $\lambda=0$ and coupled at $\lambda=1.$ The free
energy difference between these two states is defined as the excess
free energy of binding, $\Delta G_{b}$.\cite{Gallicchio2011adv} 

The binding free energy calculation protocol entails simulating the
system at series of $\lambda$ values spaced between 0 and 1 and collecting
binding energy samples at each state. The binding energy values from
each $\lambda$ state are then processed using UWHAM\cite{tan2012theory}
to obtain the excess free energy of binding $\Delta G_{b}$ and corresponding
uncertainty. The standard free energy of binding $\Delta G_{b}^{\circ}$
is obtained by adding the concentration and binding site volume term
to the excess free energy (see Computational Details).

Average interaction energies $\Delta E_{{\rm b}}$ for analysis are
obtained by averaging the binding energy values of the complexes from
the ensemble of conformations at the bound state at $\lambda=1$.
The uncertainties of binding energy values are estimated from the
standard error of the mean. The reorganization free energies for binding,
defined as $\Delta G_{{\rm reorg}}^{\circ}=\Delta G_{b}^{\circ}-\Delta E_{b}$,
are obtained from the corresponding values of the standard binding
free energy and of the binding energy. The uncertainty of the reorganization
free energy is obtained by standard error propagation.

As an alternative to simulating each alchemical $\lambda$ state independently,
to accelerate the convergence of free energy calculations, in this
work we utilize an Hamiltonian replica-exchange approach\cite{Gallicchio2008,gallicchio2015asynchronous}
where $\lambda$ values are exchanged between molecular dynamics replicas,
allowing the mixing of intermolecular degrees of freedom to explore
the conformational space efficiently.\cite{Gallicchio2008}

\subsection*{Computational Details}

\subsubsection*{Hydration Site Analysis (HSA) in explicit solvent}

Three D3 representative receptor structures were used for the Hydration
Site Analysis (HSA) in explicit solvent. The receptor structures considered
are those corresponding to the complexes of D3 with (-)-stepholidine,
C3 butyl (\textbf{1c}) and C9 butyl derivatives \cite{madapa2016synthesis}
as obtained from individual induced fit docking (IFD) simulations\cite{Sherman2006}
using the crystal structure receptor configuration of the dopamine
D3 receptor (PDB ID - 3PBL) as a starting point. The IFD protocol
was performed in five steps: generation of ligand conformations, initial
docking with reduced receptor atom van der Waal radii, side chain
minimization with Prime\cite{Jacobson:Pincus:Rapp:Day:Honig:Shaw:Friesner:2003,Jacobson2002},
a second docking step using the new receptor configuration and finally
pose scoring. Receptor-ligand configuration with the highest IFD score
ranking was selected, except in the case where the highest scored
pose did not maintain the well conserved Asp 110$^{3.32}$ salt-bridge.
The apo receptor structure from each highest scored pose, was then
used for Hydration Site Analysis (HSA).

The explicit solvent simulations for Hydration Site Analysis (HSA)
were conducted with the AMBER\cite{AmberMD2013} software package
with the OPC\cite{Izadi2014} water model with positional restraints
on all heavy atoms with a force constant of $10.0$ kcal/mol/$\mathring{{\rm A}}^{2}$.
Each system was minimized and thermalized for 2.0 ns under NPT conditions
of 1 atm and 300K. During the production run, MD simulations were
performed for $10.0$ ns under NVT conditions and snapshots of the
trajectory were collected every 1.0 ps. High density spherical regions
of 1${\rm \mathring{A}}$ radius were identified using a clustering
analysis on the water molecules which lies within 8 $\mathring{{\rm A}}$
of the superimposed ligand in D3 binding site. Individual hydration
sites were then populated with all water molecules that lies within
1 $\mathring{{\rm A}}$ of the corresponding hydration site center.
Average solvation energies were calculated for each site by calculating
the energies of the water molecules within 1 ${\rm \mathring{A}}$
of each hydration site center in all 10,000 frames of the trajectory.
For technical reasons, HSA employed a different force field (AMBER
ff14SB force field\cite{Maier2015}) than that for the binding free
energy calculations (OPLS/AA). The purpose of HSA is to obtain semi-quantitative
estimates of the energies of enclosed water molecules as well as their
locations. On a qualitative level, The large increase of binding affinities
when including enclosed hydration effects (observed below) is not
expected to depend on the choice of the force field.

\subsubsection*{System preparation for the binding free energy calculations}

The bound ligand was removed from the co-crystallized structure of
Dopamine D3 receptor with eticlopride\cite{Chien2010} along with
crystallographic waters. Protonation states were adjusted to reflect
neutral pH conditions. The receptor structure was prepared using the
Protein Preparation Wizard of the Maestro version 2016-3 (Schrodinger
Inc.). The prepared protein structure was used to generate the receptor
grid for docking using default parameters. Docking was performed with
Standard Precision (SP) version of the Glide program (Schrodinger
Release 2016-3).\cite{friesner2004glide} Positional constraints were
applied to the alkyl nitrogen of the (-)-stepholidine and all the
analogues to maintain the salt-bridge interaction with Asp 110$^{3.32}$
of the D3 receptor. The hydroxyl and thiol groups of the receptor,
such as of residues Ser 182$^{{\rm ECL2}}$, Ser 192$^{5.42}$, Ser
196$^{5.46}$, Thr 115$^{3.37}$, Thr 369$^{7.39}$, Cys 114$^{3.36}$
located near the binding site were allowed to rotate during docking. 

The (-)-stepholidine C3 analogues were built using the Maestro program
(Schrodinger Release 2016-3). Alternative protonation states as well
as chiral forms were generated for the $7\pm2$ pH range using the
LigPrep facility (Schrodinger Inc.) and ionization penalties were
calculated with Epik\cite{shelley2007epik} at pH 7. The ionization
free energies were recorded and added to the binding free energy estimates
to compute the predicted binding free energies. Only states where
the alkyl nitrogen is protonated were selected for docking calculations.
We also included in the docking study the two chiral forms of the
protonated alkyl nitrogen for each compound as generated by LigPrep
(Schrodinger Release 2016-3). 

Binding poses generated by docking were selected based on their docking
scores and presence of an ionic interaction between the protonated
alkyl nitrogen and the carboxylate group of Asp110$^{3.32}$. The
derivatives considered here are all stereoisomers with the S configuration
at the chiral carbon connecting ring B and ring C of the (-)-stepholidine
core (Table~\ref{tab:Stepholidine_derivatives_schematic}). The adjacent
protonated alkyl nitrogen atom is found always in the S configuration
while maintaining the salt-bridge interaction.

The starting conformations of complexes from docking underwent energy
minimization and thermalization. Hamiltonian Replica-exchange Molecular
dynamics simulations were performed starting from the thermalized
structures using 28 intermediate lambda states distributed as follows:
0.0, 0.002, 0.005, 0.008, 0.009, 0.01, 0.0105, 0.012, 0.0135, 0.015,
0.02, 0.0225, 0.025, 0.03, 0.035, 0.04, 0.07, 0.1, 0.25, 0.35, 0.45,
0.55, 0.65, 0.71, 0.78, 0.85, 0.92, and 1.0. The volume of the binding
site, $V_{{\rm site}}$ is defined as the spherical volume in which
the center of mass of ligand is within $3.5{\rm \;{\rm \mathring{A}}}$
of the center of mass of the binding site of the D3 receptor, defined
as the center of mass of the ${\rm C_{{\rm \alpha}}}$ atoms of the
residues 110, 111, 114, 183, 188, 346, 349 and the ${\rm C_{{\rm \beta}}}$
atoms of residues 342, 349 and the backbone nitrogen atom of residue
111. The binding site volume restraint is implemented as a flat-bottom
spherical harmonic potential with force constant of 3 kcal/mol/${\rm \mathring{A}^{2}}$
and tolerance of $3.5\:{\rm \mathring{A}}$ which resulted in a free
energy penalty $\Delta G_{{\rm t}}^{\circ}$ for transferring the
ligand from a solution of concentration $C^{\circ}$ to a volume of
size $V_{{\rm site}}$, of about 1.32 kcal/mol, calculated from the
following expression:

\begin{equation}
\Delta G_{{\rm t}}^{\circ}=-k_{B}TlnC^{\circ}V_{{\rm site}}\label{eq:c0vsite}
\end{equation}
 The receptor conformation was loosely restrained to the crystallographic
structure using flat-bottom positional restraints with a force constant
of 25 kcal/mol/${\rm {\rm \mathring{A}}^{2}}$ and a tolerance of
1.5 ${\rm {\rm \mathring{A}}}$ applied to the backbone ${\rm C_{{\rm \alpha}}}$
atoms, except for six residues 180-185 of the ECL2 loop to account
for its flexibility.

Temperature replica-exchange simulations were carried out to obtain
conformational reservoirs of the apo receptor.\cite{Gallicchio2012a}
These utilized 23 replicas distributed between 300 and 400K.\cite{Gallicchio2012a}
The conformational ensemble collected at 300K was used as a source
of apo-receptor conformations in the replica-exchange simulations.
Conformational reservoirs for each ligand were generated similarly
using 8 replicas distributed between 300 and 600K. During the simulation,
conformations of receptor and ligands were randomly selected from
the conformational reservoirs during exchanges at the fully uncoupled
state.

Single-decoupling binding free energy calculations were performed
for approximately 1 ns per replica for a total of 28 ns per complex.
Binding energies samples from the last 500 ps were used for the binding
free energy estimates. Each cycle of replica lasted 10 ps with 1 fs
MD time-step. Binding energies were collected every 10 ps. Most of
the calculations were carried out at the XSEDE SuperMIC and Stampede2
clusters utilizing CPU\textquoteright{}s and MIC devices.

To improve the convergence of the binding energies near the uncoupled
state at $\lambda=0$, we employ a soft core binding energy function
as described elsewhere.\cite{GallicchioSAMPL4,tan2012theory} The
binding energies were analyzed using the UWHAM R-statistical package\cite{tan2012theory}
to yield the binding free energy $\lyxmathsym{\textgreek{D}}G_{{\rm b}}^{\circ}$.
As mentioned, the average interaction energy $\lyxmathsym{\textgreek{D}}E_{{\rm b}}$
of each complex was obtained from the value of the average binding
energy at the coupled state ($\lambda=1)$. Reorganization free energies
$\lyxmathsym{\textgreek{D}}G_{{\rm reorg}}^{\circ}$ were measured
as the difference between the binding free energy and the average
binding energy as $\Delta G_{{\rm reorg}}^{\circ}=\Delta G_{{\rm b}}^{\circ}-\Delta E_{{\rm b}}$.

\subsubsection*{Synthesis and experimental assays of (-)-stepholidine C3 analogues}

Compounds \textbf{1a}-\textbf{1f} were synthesized using the procedure
developed as shown in Supplementary Fig.~S1. Commercially available
dihydroxy benzaldehyde, \textbf{4} was selectively protected with
a benzyl group to give compound \textbf{5}. Second, the phenolic group
of aldehyde \textbf{5} was protected with a silyl group and the intermediate
was subjected to a Henry condensation reaction to give nitrostyrene
\textbf{6}. Reduction of nitro compound \textbf{6} using ${\rm LiBH_{4}}$
yielded primary amine \textbf{7}. Aminolysis of lactone \textbf{8}
with primary amine \textbf{7} was carried out to give amide alcohol
\textbf{9}, which was acetylated to afford \textbf{10}. Ring B of
the tetrahydroprotoberberine (THPB) scaffold was formed via Bischler-Napieralski
cyclization followed by asymmetric hydrogenation using Noyori\textquoteright{}s
catalyst and formic acid/triethylamine mixture to generate \textbf{11}
with good yield (88\%). Hydrolysis of the acetyl group and subsequent
chlorination endowed us the tetracyclic scaffold of THPB in compound
\textbf{12}. The enantiomeric excess of this common precursor was
found to be 90.2\% (chiral HPLC) and it was used for further analogue
generation. Alkylation of compound \textbf{12} followed by debenzylation
provided us the C3 analogues \textbf{1a}-\textbf{1f}.

All the (-)-stepholidine C3 analogues were biochemically evaluated
by primary and secondary radioligand binding assays with the dopamine
receptor to obtain the inhibition constants of binding, $K_{i}$ and
reported in Table~\ref{tab:Measured-inhibition-constants-C3}. Both
the primary and secondary radioligand binding assays were done at
the PDSP facility (http://pdsp.med.unc.edu/). In the primary binding
assays, compounds were tested at single concentrations ($10$ \textgreek{m}M)
in quadruplicate in 96-well plates. Compounds that showed a minimum
of $50$\% inhibition at $10$ \textgreek{m}M were tagged for secondary
radioligand binding assays to determine equilibrium binding affinity
at specific targets. In the secondary binding assays, selected compounds
were tested in triplicate sets (3 sets of 96-well plates) at eleven
different concentrations out of which eight are in nanomolar range
($0.1$, $0.3$, $1$, $3$, $10$, $30$, $100$ and $300$ nM) and
rest of the three concentration in micromolar range ($1$, $3$, and
$10$ \textgreek{m}M). Both primary and secondary radioligand binding
assays were carried out in a final of volume of $125$ \textgreek{m}l
per well in appropriate binding buffer. The hot ligand concentration
was usually at a concentration close to the $K_{d}$ (unless otherwise
indicated). Total binding and nonspecific binding were determined
in the absence and presence of $10$ \textgreek{m}M Chlorpromazine,
which was used as a reference compound. In brief, plates were usually
incubated at room temperature and in the dark for $90$ min. Reactions
were stopped by vacuum filtration onto $0.3$\% polyethyleneimine
(PEI) soaked 96-well filter mats using a 96-well Filtermate harvester,
followed by three washes with cold wash buffer. Scintillation cocktail
was then melted onto the microwave-dried filters on a hot plate and
radioactivity was counted in a Microbeta counter. For detailed experimental
details, please refer to the PDSP website http://pdsp.med.unc.edu/
and click on \textquoteleft{}Binding Assay\textquoteright{} or \textquoteleft{}Functional
Assay\textquoteright{} on the menu bar.

\section*{Results}

\subsection*{Biochemical evaluation of (-)-stepholidine C3 analogues}

The inhibition constants for binding of the C3 analogues are reported
in Table. \ref{tab:Measured-inhibition-constants-C3}. The C3 analogues
showed relatively stronger inhibition of binding at the dopamine D3
receptor compared to that of C9 analogues tested previously.\cite{madapa2016synthesis}
The length of the C3 substitution has generally a small influence
on their measured affinities in this set. However, the analogues with
the longest C3 pentyl and hexyl substituent (\textbf{1d }and \textbf{1e})
exhibit a slightly stronger affinity (Table~\ref{tab:Measured-inhibition-constants-C3})

\begin{table}[h]
\caption{Measured inhibition constants of binding ($K_{i}$) for the (-)-stepholidine
C3 analogues against the dopamine D3 receptor.\label{tab:Measured-inhibition-constants-C3}}

\centering{}%
\begin{tabular}{ccc}
\hline 
{\footnotesize{}Compounds} & {\footnotesize{}C3-substituent} & \multicolumn{1}{c}{{\footnotesize{}$K_{i}^{a,b}$}}\tabularnewline
\hline 
{\footnotesize{}1a} & {\footnotesize{}Et} & {\footnotesize{}$40.0$}\tabularnewline
{\footnotesize{}1b} & {\footnotesize{}n-Pr} & {\footnotesize{}$46.0$}\tabularnewline
{\footnotesize{}1c} & {\footnotesize{}n-Bu} & {\footnotesize{}$51.0$}\tabularnewline
{\footnotesize{}1d} & {\footnotesize{}n-Pen} & {\footnotesize{}$33.0$}\tabularnewline
{\footnotesize{}1e} & {\footnotesize{}n-Hex} & {\footnotesize{}$26.0$}\tabularnewline
{\footnotesize{}1f} & {\footnotesize{}2-fluroethyl} & {\footnotesize{}$86.0$}\tabularnewline
\hline 
\end{tabular}\\
\begin{minipage}[t]{1\columnwidth}%
\begin{flushleft}
{\scriptsize{}$^{a}$ In nM. Experiments were carried out in triplicate
- uncertainties are estimated as 13\% of reported $K_{i}$; $^{b}$${\rm [^{3}H]}$
N-methylspiperone used as radioligand; chlorpromazine used as a reference
compound with $K_{i}=11.0$ nM. The biochemical details of the assay
are provided in the main text.}
\par\end{flushleft}%
\end{minipage}
\end{table}

\subsection*{Binding Free Energy Calculations}

We employed the enclosed hydration model described above to study
six derivatives of (-)-stepholidine substituted at the C3 position
with and without the enclosed hydration corrections to probe the effects
of enclosed hydration on the binding free energy predictions (Table~\ref{tab:C3-results}).

\begin{figure}
\begin{centering}
\includegraphics[scale=0.2]{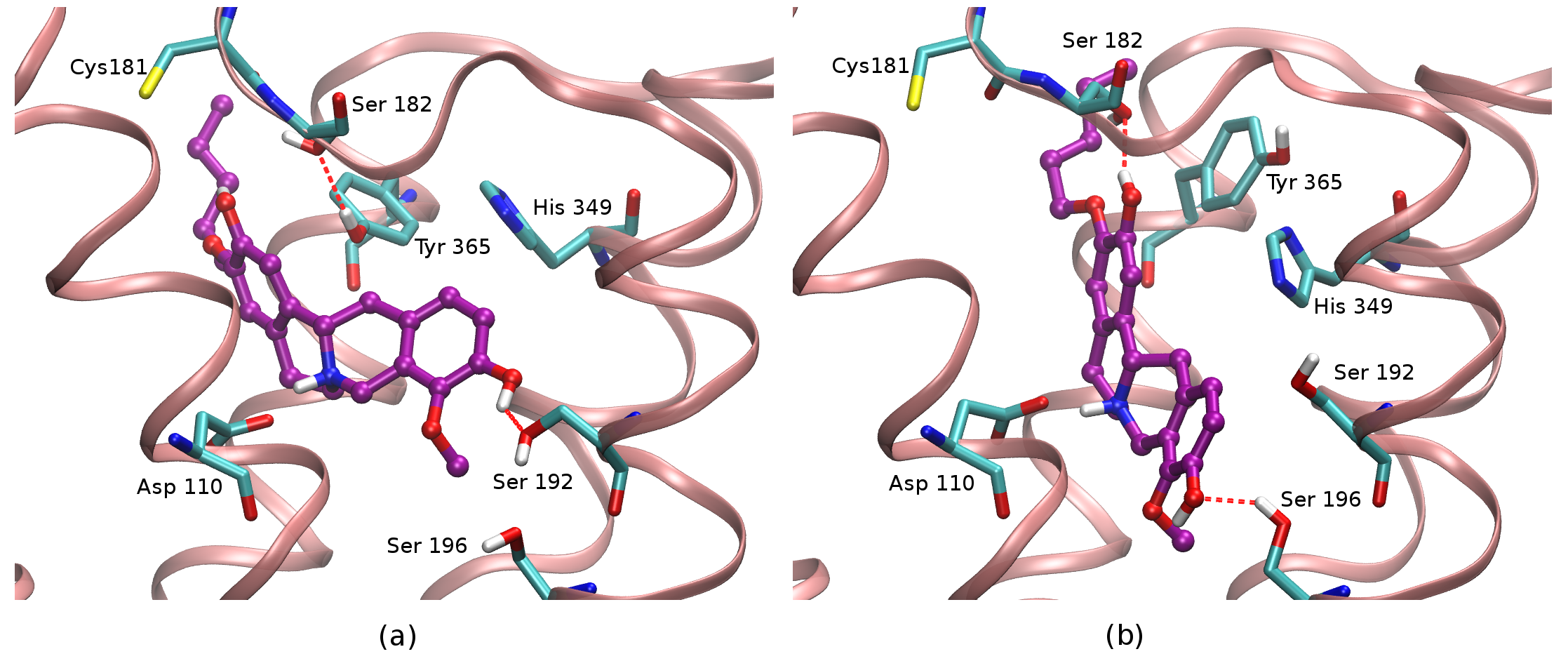}
\par\end{centering}

\caption{a) The C3 pentyl analogue (3e, purple) of (-)-stepholidine is observed
to interact with Ser 192 of the receptor at the orthosteric binding
site. In order for the C3 analogues to interact with Ser 192, the
C10 hydroxyl group is placed in proximity of Ser 192; b) The 3e C3
analogue in another observed binding pose in which it interacts with
Ser 196, rather than Ser 192. In this pose, ring D of the (-)-stepholidine
core is bound deeper into the orthosteric binding site and the ligand
is twisted causing Tyr 365 in the SBS to rotate and move away from
Ser 182 of ECL2. The receptor is represented as a pink ribbon. \label{fig:C9-C3_comparison}}
\end{figure}

The (-)-stepholidine C3 analogues are substituted at the third position
of ring A of the (-)-stepholidine core. To accommodate the long alkyl
chain substituents, the C3 analogues (Fig.~\ref{fig:C9-C3_comparison})
are found to dock to the dopamine D3 receptor in a binding pose so
that the alkyl chain occupies the secondary binding site (SBS). This
has the important consequence that ring D, occupies the OBS so to
maintain the salt bridge with Asp 110$^{3.32}$ in contrast to C9
analogues where ring A occupy the OBS\cite{madapa2016synthesis}.

The enclosed hydration model is found to be an essential ingredient
to reproduce the observed affinities. Binding free energy estimates
of C3 derivatives obtained without enclosed hydration grossly underestimate
the magnitudes of the experimental affinities derived from the measured
inhibition constants of binding (Table~\ref{tab:C3-results}, 2nd
and 3rd columns). In contrast, binding free energy calculated with
the enclosed hydration model are significantly more favorable and
substantially in better quantitative agreement with the experiments
than without enclosed hydration (Table~\ref{tab:C3-results}, 2nd
and 6th columns). When employing the enclosed hydration model, the
root mean square error (RMSE) is reduced by a factor of 6 and, while
variations in the experimental values are slight (Table~\ref{tab:C3-results},
2nd column), the level of correlation increased from less than zero
to $64$\%. The values of the calculated binding free energies with
enclosed hydration are all within $2$ kcal/mol of the experiments. 

\begin{table}[h]
\caption{Experimental and calculated binding free energies, average binding
energies and reorganization free energies of the (-)-stepholidine
C3 analogues with and without enclosed hydration corrections. \label{tab:C3-results}}

\raggedright{}{\footnotesize{}}%
\begin{tabular*}{0.8\paperwidth}{@{\extracolsep{\fill}}lcccccccc}
\toprule 
{\footnotesize{}Compound} & {\footnotesize{}$\Delta G_{{\rm exp}}^{\circ^{a,b}}$} & \multicolumn{3}{c}{{\footnotesize{}Without enclosed hydration model}} &  & \multicolumn{3}{c}{{\footnotesize{}With enclosed hydration model}}\tabularnewline
\cmidrule{3-5} \cmidrule{7-9} 
 &  & {\footnotesize{}$\Delta G_{{\rm calc}}^{\circ^{b,c}}$} & {\footnotesize{}$\Delta E_{{\rm b}}^{^{b,c}}$} & {\footnotesize{}$\Delta G_{{\rm reorg}}^{\circ^{b,c}}$} &  & {\footnotesize{}$\Delta G_{{\rm calc}}^{\circ^{b,c}}$} & {\footnotesize{}$\Delta E_{{\rm b}}^{^{b,c}}$} & {\footnotesize{}$\Delta G_{{\rm reorg}}^{\circ^{b,c}}$}\tabularnewline
\midrule
{\footnotesize{}1a} & {\footnotesize{}$-10.1$} & {\footnotesize{}$-2.2$} & {\footnotesize{}$-36.9$} & {\footnotesize{}$34.7$} &  & {\footnotesize{}$-8.8$} & {\footnotesize{}$-42.5$} & {\footnotesize{}$33.7$}\tabularnewline
{\footnotesize{}1b} & {\footnotesize{}$-10.0$} & {\footnotesize{}$-2.3$} & {\footnotesize{}$-38.0$} & {\footnotesize{}$35.7$} &  & {\footnotesize{}$-10.4$} & {\footnotesize{}$-44.7$} & {\footnotesize{}$34.3$}\tabularnewline
{\footnotesize{}1c} & {\footnotesize{}$-10.0$} & {\footnotesize{}$-1.8$} & {\footnotesize{}$-40.3$} & {\footnotesize{}$38.5$} &  & {\footnotesize{}$-11.5$} & {\footnotesize{}$-48.1$} & {\footnotesize{}$36.6$}\tabularnewline
{\footnotesize{}1d} & {\footnotesize{}$-10.2$} & {\footnotesize{}$-0.3$} & {\footnotesize{}$-43.7$} & {\footnotesize{}$43.4$} &  & {\footnotesize{}$-10.6$} & {\footnotesize{}$-55.6$} & {\footnotesize{}$45.0$}\tabularnewline
{\footnotesize{}1e} & {\footnotesize{}$-10.4$} & {\footnotesize{}$-3.9$} & {\footnotesize{}$-39.6$} & {\footnotesize{}$35.7$} &  & {\footnotesize{}$-12.5$} & {\footnotesize{}$-55.2$} & {\footnotesize{}$42.7$}\tabularnewline
{\footnotesize{}1f} & {\footnotesize{}$-9.6$} & {\footnotesize{}$-3.1$} & {\footnotesize{}$-32.7$} & {\footnotesize{}$29.6$} &  & {\footnotesize{}$-8.9$} & {\footnotesize{}$-43.2$} & {\footnotesize{}$34.3$}\tabularnewline
\midrule 
{\footnotesize{}RMSE$^{b,d}$} &  & {\footnotesize{}$7.9$} &  &  &  & {\footnotesize{}$1.2$} &  & \tabularnewline
\multicolumn{1}{l}{{\footnotesize{}Correlation coefficient (r)}} &  & {\footnotesize{}$-0.014$} &  &  &  & {\footnotesize{}$0.64$} &  & \tabularnewline
\bottomrule
\end{tabular*}\\
\begin{minipage}[t]{1\columnwidth}%
\begin{flushleft}
{\scriptsize{}$^{a}$ Experimental affinities are calculated using
the relation $\Delta G_{{\rm exp}}^{\circ}=k_{B}T\, lnK_{i}$ where
$K_{i}$ is the inhibition constant of binding, $k_{B}$is the Boltzmann's
constant. $^{b}$In kcal/mol. $^{c}$Approximate uncertainties for
all measurements are implied by the number of significant figures;
the actual values of the uncertainties for each measurement are provided
in Supplementary Table ST3. $^{d}$Root mean square error relative
to the experimental binding free energies.}
\par\end{flushleft}%
\end{minipage}
\end{table}

\section*{Discussion}

Though efficient and faster convergence of binding free energy calculations
can be achieved using implicit solvent models, these lack the ability
to model solvent heterogeneity and confinement in molecular simulations,
especially within deep protein binding pockets. In absence of ligand,
enclosed water molecules form network of interactions among themselves
and with receptor atoms, which are fundamentally different from those
present in the bulk and solvent exposed regions of the protein.\cite{Young2007,Haider2016hsa}
Water molecules which maintain favorable contacts with the protein
or act as bridging waters generally disfavor binding when displaced
by the ligand. However, enegetically and entropically frustrated water
molecules such as those trapped within the hydrophobic regions of
the binding site, favor binding when displaced by the ligand. In this
work, we have employed for the first time a hybrid computational model
involving explicit and implicit solvation to include the thermodynamics
of confined water in the calculation of the binding free energies
of protein-ligand complexes. We applied the model to calculate the
binding free energies for a series of novel compounds as potential
ligands of the dopamine D3 receptor, which have been synthesized and
assayed for activity as part of this work. In all cases tested, binding
free energies were observed to be more favorable in the presence of
enclosed hydration effects compared to the conventional implicit solvent
model. The enhancement of binding affinities with the enclosed hydration
model is in accord with the idea that energetically frustrated enclosed
water molecules contributed favorably to binding when displaced by
the ligand.

In this study, we identify a class of dopamine D3 receptor ligands
which are more powerful than those previously synthesized and assayed.\cite{madapa2016synthesis}
The affinities of the (-)-stepholidine C3 analogues, synthesized in
this work, justifies the motivation of synthesis to increase interaction
at the secondary binding site (SBS) by adding substituents at the
C3 position, with the strongest affinitiy being observed for the longest
substitution (\textbf{1e}) in agreement with the computational predictions
(Table~\ref{tab:Measured-inhibition-constants-C3}). The modeling
approach introduced here has provided key insights for this system.
All of the compounds analyzed consistently maintained an ionic interaction
between the protonated alkyl nitrogen of the (-)-stepholidine core
and the carboxylate group of Asp110$^{3.32}$ of the D3 receptor. 

\begin{figure}
\begin{centering}
\includegraphics[scale=0.25]{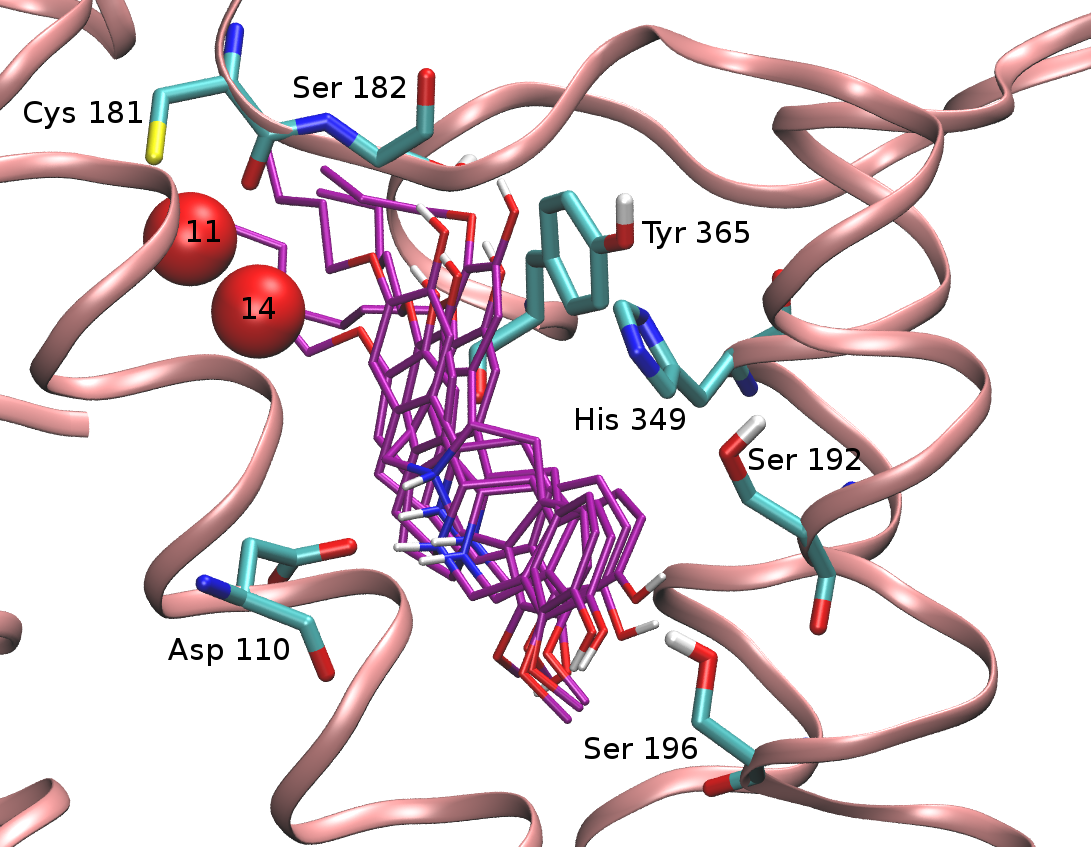}
\par\end{centering}

\caption{Representative bound poses of (-)-stepholidine C3 analogues (purple)
interacting with Ser 196 at the orthosteric binding site of the dopamine
D3 receptor is observed to displace fewer enclosed water molecules,
especially at the secondary binding site. AGBPN2 sites 11 and 14 are
not displaced in this conformation of the C3 analogues (Table~\ref{tab:HSA-score}).\label{fig:C3_watersites}}
\end{figure}

The positioning of C3 analogues within the binding site affect not
only the pattern of ligand-receptor interactions in the secondary
binding site, but crucially, also the interactions within the orthosteric
pocket as well as the pattern of displacement of energetically unfavorable
water molecules (Fig.~\ref{fig:C3_watersites}). These energetic
and structural features are ultimately reflected in the differences
of binding affinities with and without enclosed hydration effects
(Table~\ref{tab:C3-results}). When not considering enclosed hydration
effects, the calculated binding affinities of the C3 analogues are
observed to be very overly unfavorable. Inclusion of the enclosed
hydration effects in the calculation, made the calculated binding
free energies more favorable and improved the agreement with the experimental
values (Table~\ref{tab:C3-results}). 

In our model, ring D of the (-)-stepholidine C3 analogues is placed
into the orthosteric binding pocket where it is observed to interact
with Ser 192$^{5.42}$ through one hydrogen bond interaction with
the hydroxyl group at position C10. In addition, the hydrogen bond
interaction of C3 analogues is not stably maintained throughout the
simulation, as it is seen to periodically switch to an alternate hydrogen
bonding interaction with Ser 196$^{5.46}$ slightly deeper into the
orthosteric binding site (Fig.~\ref{fig:C9-C3_comparison}b). Also,
the binding of C3 analogues is observed to displace almost all enclosed
water molecules within the orthosteric binding site by placing the
(-)-stepholidine core. However, while interacting with Ser 196$^{5.46}$,
the alkoxy substituent chain at the secondary binding site (SBS) displaced
fewer enclosed water molecules. These enclosed water sites, however,
impose less energetic penalties, totaling to less than $1.5$ kcal/mol
(sites 11 and 14, see Table~\ref{tab:HSA-score} and Fig.~\ref{fig:C3_watersites}),
thereby contributing to little difference in the calculated binding
affinities between the C3 derivatives. Another interesting observation
in this pose is the displacement of Tyr 365$^{7.35}$ of helix VII
away from the secondary binding site (Fig.~\ref{fig:C9-C3_comparison}b)
and the concurrent disruption of the hydrogen bond interaction with
Ser 182$^{{\rm ECL2}}$ which stabilizes the extracellular loop 2
(ECL2) in the SBS. 

Conformational changes within the binding site may change the number
and pattern of ligand-receptor interactions\cite{GPCRFilizola2011}
as well as the hydration structure, which we know to be very sensitive
to the placement of receptor atoms. All calculations were done in
absence of the description of the cellular membrane while limiting
large backbone motions. Despite these limitations, our computational
protocol was able to correctly predict the affinities of the C3 analogues
with reasonable accuracy. 

All these observations illustrates the complexities associated with
binding of the (-)-stepholidine analogues to the dopamine D3 receptor.
They also underscore the challenges encountered in the design of effective
and selective D3 ligands/antagonists.\cite{molecularDeterminantD32012newman,medicationTargets2010D3,maramai2016dopamine,madapa2016synthesis,heidbreder2010current,Michino2017}
One major challenge is the effect of the specific remodeling of the
receptor binding site induced by ligands. In our study, induced fit
docking calculations have not revealed major structural changes for
different (-)-stepholidine analogues, although Hydration Site analysis
(HSA) revealed more significant changes in the hydration energies
and location of the hydration sites. The modeled binding affinities
of the C3 analogues in this work may reflect the limitations imposed
by the initial receptor structure. Another computational challenge
in this work has been the appropriate representation of the enclosed
hydration sites by exploiting the available topologies afforded by
the current AGBNP2 implicit solvent model.

\section*{Conclusion}

In this study, we exploited the energetics of confined water molecules
as obtained from explicit solvent simulations, and trained an implicit
solvent model to account their effects on protein-ligand binding free
energies, using a hybrid approach which proved useful for host-guest
binding thermodynamics.\cite{Pal2017} 

Protein binding sites are much more complex than host-guest systems
both in terms of structure and conformational variability. This is
the first report of the implementation of a hybrid explicit-implicit
solvent approach to calculate the binding affinities of protein ligand
complexes and its application to a series of complexes of the dopamine
D3 receptor. As we have illustrated, it is very challenging to model
with high confidence the thermodynamics of enclosed water molecules
in protein binding sites. While more research is needed to improve
and automate model parameterization and model accuracy, this study
confirms that it is both useful and viable to include enclosed hydration
effects in binding free energy calculations with implicit solvation
as an alternative to explicit modeling, which is more affected by
slow equilibration.\cite{Deng2009,BruceMacdonald2018,Clark2009a}

The experimental dissociation constants and the computational modeling
work have provided valuable insights for the design of stronger and
specific ligands of the dopamine D3 receptor. This study emphasizes
the benefits of interdisciplinary approaches by tackling difficult
rational drug design problems from different experimental, synthetic
and modeling sides.

\section*{Acknowledgements}

This work was supported in part by an Interdisciplinary Research Grant
from the City University of New York (CIRG 2313, to E.G., L.W., T.K.,
and W.W.H.). R.K.P and E.G. acknowledge support from the National
Science Foundation (NSF CAREER 1750511). E.G. acknowledges support
from the CUNY PSC program (PSC-CUNY 61211-00 49) and Levy-Kosminsky
Professorship in Physical Chemistry at Brooklyn College. T.K. and
S.R. acknowledge support from NIH SCORE grant SC3-GM095417. Molecular
simulations were conducted on the WEB computational grid at Brooklyn
College of the City University of New York, SuperMIC cluster at the
Louisiana State University High Performance Computing Center and the
Stampede II supercomputer cluster at the Texas Advanced Computing
Center supported by NSF XSEDE award TG-MCB150001. $K_{i}$ determinations
and receptor binding profiles were generously provided by the National
Institute of Mental Health\textquoteright{}s Psychoactive Drug Screening
Program, Contract \# HHSN-271-2008-00025-C (NIMH PDSP). The NIMH PDSP
is directed by Bryan L. Roth MD, PhD, at the University of North Carolina
at Chapel Hill and Project Officer Jamie Driscol at NIMH, Bethesda
MD, USA.

\section*{Author Contribution}

\textbf{Conceptualization:} Emilio Gallicchio, Tom Kurtzman\\
\textbf{Data curation: }Rajat Kumar Pal, Lauren Wickstrom\\
\textbf{Formal analysis:} Rajat Kumar Pal, Emilio Gallicchio\\
\textbf{Investigation:} Rajat Kumar Pal, Satishkumar Gadhiya, Steve
Ramsey, Lauren Wickstrom, Emilio Gallicchio\\
\textbf{Methodology:} Rajat Kumar Pal, Emilio Gallicchio, Steve Ramsey,
Pierpaolo Cordone, Tom Kurtzman, Wayne W. Harding\\
\textbf{Software:} Rajat Kumar Pal, Emilio Gallicchio, Tom Kurtzman\\
\textbf{Supervision:} Emilio Gallicchio, Tom Kurtzman, Wayne W. Harding,
Lauren Wickstrom\\
\textbf{Validation:} Rajat Kumar Pal, Satishkumar Gadhiya \\
\textbf{Writing \textendash{} original draft:} Rajat Kumar Pal, Emilio
Gallicchio, Satishkumar Gadhiya \\
\textbf{Writing \textendash{} review \& editing}: Everyone\\

\pagebreak{}

\bibliographystyle{plos2015}

\begin{thebibliography}{10}

\bibitem{de2010role}
de~Beer S, Vermeulen N, Oostenbrink C.
\newblock {The Role of Water Molecules in Computational Drug Design}.
\newblock Curr Top Med Chem. 2010;10(1):55--66.
\newblock doi:{10.2174/156802610790232288}.

\bibitem{Li2007}
Li Z, Lazaridis T.
\newblock {Water at biomolecular binding interfaces}.
\newblock Phys Chem Chem Phys. 2007;9(5):573--581.
\newblock doi:{10.1039/b612449f}.

\bibitem{mancera2007molecular}
Mancera RL.
\newblock {Molecular modeling of hydration in drug design.}
\newblock Curr Opin Drug Discov Devel. 2007;10(3):275--280.

\bibitem{ball2008water}
Ball P.
\newblock {Water as an Active Constituent in Cell Biology Water as an Active
  Constituent in Cell Biology}.
\newblock Chem Rev. 2008;108(1):74--108.
\newblock doi:{10.1021/cr068037a}.

\bibitem{ladbury1996just}
Ladbury JE.
\newblock {Just add water! The effect of water on the specificity of
  protein-ligand binding sites and its potential application to drug design}.
\newblock Chem Biol. 1996;3(12):973--980.
\newblock doi:{10.1016/S1074-5521(96)90164-7}.

\bibitem{nguyen2014thermodynamics}
Nguyen CN, Cruz A, Gilson MK, Kurtzman T.
\newblock {Thermodynamics of water in an enzyme active site: Grid-based
  hydration analysis of coagulation factor xa}.
\newblock J Chem Theory Comput. 2014;10(7):2769--2780.
\newblock doi:{10.1021/ct401110x}.

\bibitem{Setny2010}
Setny P, Baron R, McCammon JA.
\newblock {How can hydrophobic association be enthalpy driven?}
\newblock J Chem Theory Comput. 2010;6(9):2866--2871.
\newblock doi:{10.1021/ct1003077}.

\bibitem{Haider2016hsa}
Haider K, Wickstrom L, Ramsey S, Gilson MK, Kurtzman T.
\newblock {Enthalpic Breakdown of Water Structure on Protein Active-Site
  Surfaces}.
\newblock J Phys Chem B. 2016;120(34):8743--8756.
\newblock doi:{10.1021/acs.jpcb.6b01094}.

\bibitem{Pal2017}
Pal RK, Haider K, Kaur D, Flynn W, Xia J, Levy RM, et~al.
\newblock {A combined treatment of hydration and dynamical effects for the
  modeling of host-guest binding thermodynamics: the SAMPL5 blinded challenge}.
\newblock J Comput Aided Mol Des. 2017;31(1):29--44.
\newblock doi:{10.1007/s10822-016-9987-z}.

\bibitem{Beuming2009}
Beuming T, Farid R, Sherman W.
\newblock {High-energy water sites determine peptide binding affinity and
  specificity of PDZ domains}.
\newblock Protein Sci. 2009;18(8):1609--1619.
\newblock doi:{10.1002/pro.177}.

\bibitem{Young2007}
Young T, Abel R, Kim B, Berne BJ, Friesner RA.
\newblock {Motifs for molecular recognition exploiting hydrophobic enclosure in
  protein-ligand binding}.
\newblock Proc Natl Acad Sci. 2007;104(3):808--813.
\newblock doi:{10.1073/pnas.0610202104}.

\bibitem{huggins2012application}
Huggins DJ.
\newblock {Application of inhomogeneous fluid solvation theory to model the
  distribution and thermodynamics of water molecules around biomolecules}.
\newblock Phys Chem Chem Phys. 2012;14(43):15106--15117.
\newblock doi:{10.1039/c2cp42631e}.

\bibitem{ross2012rapid}
Ross GA, Morris GM, Biggin PC.
\newblock {Rapid and Accurate Prediction and Scoring of Water Molecules in
  Protein Binding Sites}.
\newblock PLoS One. 2012;7(3):e32036.
\newblock doi:{10.1371/journal.pone.0032036}.

\bibitem{Ross2015}
Ross GA, Bodnarchuk MS, Essex JW.
\newblock {Water Sites, Networks, And Free Energies with Grand Canonical Monte
  Carlo}.
\newblock J Am Chem Soc. 2015;137(47):14930--14943.
\newblock doi:{10.1021/jacs.5b07940}.

\bibitem{bodnarchuk2014strategies}
Bodnarchuk MS, Viner R, Michel J, Essex JW.
\newblock {Strategies to Calculate Water Binding Free Energies in Protein -
  Ligand Complexes}.
\newblock J Chem Inf Model. 2014;54(6):1623--1633.

\bibitem{sindhikara2013analysis}
Sindhikara DJ, Hirata F.
\newblock {Analysis of biomolecular solvation sites by 3D-RISM theory}.
\newblock J Phys Chem B. 2013;117(22):6718--6723.
\newblock doi:{10.1021/jp4046116}.

\bibitem{Graham2017}
Graham SE, Smith RD, Carlson HA.
\newblock {Predicting Displaceable Water Sites Using Mixed-Solvent Molecular
  Dynamics}.
\newblock J Chem Inf Model. 2018;58(2):305--314.
\newblock doi:{10.1021/acs.jcim.7b00268}.

\bibitem{Michel2009a}
Michel J, Tirado-Rives J, Jorgensen WL.
\newblock {Prediction of the water content in protein binding sites}.
\newblock J Phys Chem B. 2009;113(40):13337--13346.
\newblock doi:{10.1021/jp9047456}.

\bibitem{Macdonald}
{Bruce Macdonald} HE, Cave-Ayland C, Ross GA, Essex JW.
\newblock {Ligand Binding Free Energies with Adaptive Water Networks:
  Two-Dimensional Grand Canonical Alchemical Perturbations}.
\newblock J Chem Theory Comput. 2018;14(12):6586--6597.
\newblock doi:{10.1021/acs.jctc.8b00614}.

\bibitem{heidbreder2010current}
Heidbreder CA, Newman AH.
\newblock {Current perspectives on selective dopamine D3 receptor antagonists
  as pharmacotherapeutics for addictions and related disorders}.
\newblock Ann N Y Acad Sci. 2010;1187(1):4--34.
\newblock doi:{10.1111/j.1749-6632.2009.05149.x}.

\bibitem{Gadhiya2018}
Gadhiya S, Cordone P, Pal RK, Gallicchio E, Wickstrom L, Kurtzman T, et~al.
\newblock {New Dopamine D3-Selective Receptor Ligands Containing a
  6-Methoxy-1,2,3,4-tetrahydroisoquinolin-7-ol Motif}.
\newblock ACS Med Chem Lett. 2018;9(10):990--995.
\newblock doi:{10.1021/acsmedchemlett.8b00229}.

\bibitem{madapa2016synthesis}
Madapa S, Gadhiya S, Kurtzman T, Alberts IL, Ramsey S, Reith M, et~al.
\newblock {Synthesis and evaluation of C9 alkoxy analogues of (-)-stepholidine
  as dopamine receptor ligands}.
\newblock Eur J Med Chem. 2017;125:255--268.
\newblock doi:{10.1016/j.ejmech.2016.09.036}.

\bibitem{GPCRwaterPathway2014}
Yuan S, Filipek S, Palczewski K, Vogel H.
\newblock {Activation of G-protein-coupled receptors correlates with the
  formation of a continuous internal water pathway}.
\newblock Nat Commun. 2014;5(1):4733.
\newblock doi:{10.1038/ncomms5733}.

\bibitem{maramai2016dopamine}
Maramai S, Gemma S, Brogi S, Campiani G, Butini S, Stark H, et~al.
\newblock {Dopamine D3 Receptor Antagonists as Potential Therapeutics for the
  Treatment of Neurological Diseases}.
\newblock Front Neurosci. 2016;10(OCT):451.
\newblock doi:{10.3389/fnins.2016.00451}.

\bibitem{Volkow2007}
Volkow ND, Fowler JS, Wang GJ, Swanson JM, Telang F.
\newblock {Dopamine in drug abuse and addiction: Results of imaging studies and
  treatment implications}.
\newblock Arch Neurol. 2007;64(11):1575--1579.
\newblock doi:{10.1001/archneur.64.11.1575}.

\bibitem{Brooks2000}
Brooks DJ.
\newblock {Dopamine agonists: Their role in the treatment of Parkinson's
  disease}.
\newblock J Neurol Neurosurg Psychiatry. 2000;68(6):685--689.
\newblock doi:{10.1136/jnnp.68.6.685}.

\bibitem{Chien2010}
Chien EYT, Liu W, Zhao Q, Katritch V, {Won Han} G, Hanson MA, et~al.
\newblock {Structure of the Human Dopamine D3 Receptor in Complex with a D2/D3
  Selective Antagonist}.
\newblock Science. 2010;330(6007):1091--1095.
\newblock doi:{10.1126/science.1197410}.

\bibitem{cho2010current}
Cho DI, Zheng M, Kim KM.
\newblock {Current perspectives on the selective regulation of dopamine D2 and
  D3 receptors}.
\newblock Arch Pharm Res. 2010;33(10):1521--1538.
\newblock doi:{10.1007/s12272-010-1005-8}.

\bibitem{InsightsD1Ragonist2012li}
Li B, Li W, Du P, Yu KQ, Fu W.
\newblock {Molecular insights into the D1R agonist and D2R/D3R antagonist
  effects of the natural product (-)-stepholidine: Molecular modeling and
  dynamics Simulations}.
\newblock J Phys Chem B. 2012;116(28):8121--8130.
\newblock doi:{10.1021/jp3049235}.

\bibitem{stepholidine2007clinicalMo}
Mo J, Guo Y, Yang YS, Shen JS, Jin GZ, Zhen X.
\newblock {Recent Developments in Studies of l-Stepholidine and its Analogs:
  Chemistry, Pharmacology and Clinical Implications}.
\newblock Curr Med Chem. 2007;14(28):2996--3002.
\newblock doi:{10.2174/092986707782794050}.

\bibitem{extrapyramidal1997knable}
Knable MB, Heinz A, Raedler T, Weinberger DR.
\newblock {Extrapyramidal side effects with risperidone and haloperidol at
  comparable D2 receptor occupancy levels}.
\newblock Psychiatry Res - Neuroimaging. 1997;75(2):91--101.
\newblock doi:{10.1016/S0925-4927(97)00023-1}.

\bibitem{extrapyramidal2017D2skyes}
Sykes DA, Moore H, Stott L, Holliday N, Javitch JA, Lane JR, et~al.
\newblock {Extrapyramidal side effects of antipsychotics are linked to their
  association kinetics at dopamine D2 receptors}.
\newblock Nat Commun. 2017;8(1):763.
\newblock doi:{10.1038/s41467-017-00716-z}.

\bibitem{song2014blockade}
Song R, Bi GH, Zhang HY, Yang RF, Gardner EL, Li J, et~al.
\newblock {Blockade of D3 receptors by YQA14 inhibits cocaine's rewarding
  effects and relapse to drug-seeking behavior in rats}.
\newblock Neuropharmacology. 2014;77:398--405.
\newblock doi:{10.1016/j.neuropharm.2013.10.010}.

\bibitem{medicationTargets2010D3}
Keck TM, John WS, Czoty PW, Nader MA, Newman AH.
\newblock {Identifying Medication Targets for Psychostimulant Addiction:
  Unraveling the Dopamine D3 Receptor Hypothesis}.
\newblock J Med Chem. 2015;58(14):5361--5380.
\newblock doi:{10.1021/jm501512b}.

\bibitem{D3crypticpocket2017ferruz}
Ferruz N, Doerr S, Vanase-Frawley MA, Zou Y, Chen X, Marr ES, et~al.
\newblock {Dopamine D3 receptor antagonist reveals a cryptic pocket in
  aminergic GPCRs}.
\newblock Sci Rep. 2018;8(1):1--10.
\newblock doi:{10.1038/s41598-018-19345-7}.

\bibitem{molecularDeterminantD32012newman}
Newman AH, Beuming T, Banala AK, Donthamsetti P, Pongetti K, Labounty A, et~al.
\newblock {Molecular determinants of selectivity and efficacy at the dopamine
  D3 receptor}.
\newblock J Med Chem. 2012;55(15):6689--6699.
\newblock doi:{10.1021/jm300482h}.

\bibitem{stepholidine2015panDopamine}
Meade JA, Free RB, Miller NR, Chun LS, Doyle TB, Moritz AE, et~al.
\newblock {(-)-Stepholidine is a potent pan-dopamine receptor antagonist of
  both G protein- and $\beta$-arrestin-mediated signaling}.
\newblock Psychopharmacology (Berl). 2015;232(5):917--930.
\newblock doi:{10.1007/s00213-014-3726-8}.

\bibitem{stepholidine2017mtorBzhang}
Zhang B, Guo F, Ma Y, Song Y, Lin R, Shen FY, et~al.
\newblock {Activation of D1R/PKA/mTOR signaling cascade in medial prefrontal
  cortex underlying the antidepressant effects of l-SPD}.
\newblock Sci Rep. 2017;7(1):3809.
\newblock doi:{10.1038/s41598-017-03680-2}.

\bibitem{stepholidine2018manuszak}
Manuszak M, Harding W, Gadhiya S, Ranaldi R.
\newblock {(-)-Stepholidine reduces cue-induced reinstatement of cocaine
  seeking and cocaine self-administration in rats}.
\newblock Drug Alcohol Depend. 2018;189:49--54.
\newblock doi:{10.1016/j.drugalcdep.2018.04.030}.

\bibitem{D1D2agonist2007Fu}
Fu W, Shen J, Luo X, Zhu W, Cheng J, Yu K, et~al.
\newblock {Dopamine D1 receptor agonist and D2 receptor antagonist effects of
  the natural product (2)-stepholidine: Molecular modeling and dynamics
  simulations}.
\newblock Biophys J. 2007;93(5):1431--1441.
\newblock doi:{10.1529/biophysj.106.088500}.

\bibitem{Lazaridis:98}
Lazaridis T.
\newblock {Inhomogeneous Fluid Approach to Solvation Thermodynamics. 1.
  Theory}.
\newblock J Phys Chem B. 1998;102(18):3531--3541.
\newblock doi:{10.1021/jp9723574}.

\bibitem{Izadi2014}
Izadi S, Anandakrishnan R, Onufriev AV.
\newblock {Building water models: A different approach}.
\newblock J Phys Chem Lett. 2014;5(21):3863--3871.
\newblock doi:{10.1021/jz501780a}.

\bibitem{nguyen2012grid}
Nguyen CN, {Kurtzman Young} T, Gilson MK.
\newblock {Grid inhomogeneous solvation theory: Hydration structure and
  thermodynamics of the miniature receptor cucurbit[7]uril}.
\newblock J Chem Phys. 2012;137(4):973--980.
\newblock doi:{10.1063/1.4733951}.

\bibitem{Oroguchi2016}
Oroguchi T, Nakasako M.
\newblock {Changes in hydration structure are necessary for collective motions
  of a multi-domain protein}.
\newblock Sci Rep. 2016;6(1):26302.
\newblock doi:{10.1038/srep26302}.

\bibitem{Gallicchio2009}
Gallicchio E, Paris K, Levy RM.
\newblock {The AGBNP2 implicit solvation model}.
\newblock J Chem Theory Comput. 2009;5(9):2544--2564.
\newblock doi:{10.1021/ct900234u}.

\bibitem{Kaminski:2001}
Kaminski GA, Friesner RA, Tirado-Rives J, Jorgensen WL.
\newblock {Evaluation and reparametrization of the OPLS-AA force field for
  proteins via comparison with accurate quantum chemical calculations on
  peptides}.
\newblock J Phys Chem B. 2001;105(28):6474--6487.
\newblock doi:{10.1021/jp003919d}.

\bibitem{Jacobson:Kaminski:Friesner:Rapp:2002}
Jacobson MP, Kaminski GA, Friesner RA, Rapp CS.
\newblock {Force field validation using protein side chain prediction}.
\newblock J Phys Chem B. 2002;106(44):11673--11680.
\newblock doi:{10.1021/jp021564n}.

\bibitem{Qiu:Shenkin:Hollinger:Still:97}
Qiu D, Shenkin PS, Hollinger FP, Still WC.
\newblock {The GB/SA continuum model for solvation. A fast analytical method
  for the calculation of approximate Born radii}.
\newblock J Phys Chem A. 1997;101(16):3005--3014.
\newblock doi:{10.1021/jp961992r}.

\bibitem{Hawkins:Cramer:Truhlar:96}
Hawkins GD, Cramer CJ, Truhlar DG.
\newblock {Parametrized models of aqueous free energies of solvation based on
  pairwise descreening of solute atomic charges from a dielectric medium}.
\newblock J Phys Chem. 1996;100(51):19824--19839.
\newblock doi:{10.1021/jp961710n}.

\bibitem{Gallicchio2010}
Gallicchio E, Lapelosa M, Levy RM.
\newblock {Binding Energy Distribution Analysis Method (BEDAM) for Estimation
  of Protein-Ligand Binding Affinities}.
\newblock J Chem Theory Comput. 2010;6:2961--2977.
\newblock doi:{10.1021/ct1002913}.

\bibitem{Gallicchio2011adv}
Gallicchio E, Levy RM.
\newblock {Recent theoretical and computational advances for modeling
  protein-ligand binding affinities}.
\newblock Adv Protein Chem Struct Biol. 2011;85(1):27--80.
\newblock doi:{10.1016/B978-0-12-386485-7.00002-8}.

\bibitem{tan2012theory}
Tan Z, Gallicchio E, Lapelosa M, Levy RM.
\newblock {Theory of binless multi-state free energy estimation with
  applications to protein-ligand binding}.
\newblock J Chem Phys. 2012;136(14):144102.
\newblock doi:{10.1063/1.3701175}.

\bibitem{Gallicchio2008}
Gallicchio E, Levy RM, Parashar M.
\newblock {Asynchronous replica exchange for molecular simulations}.
\newblock J Comput Chem. 2008;29(5):788--794.
\newblock doi:{10.1002/jcc.20839}.

\bibitem{gallicchio2015asynchronous}
Gallicchio E, Xia J, Flynn WF, Zhang B, Samlalsingh S, Mentes A, et~al.
\newblock {Asynchronous replica exchange software for grid and heterogeneous
  computing}.
\newblock Comput Phys Commun. 2015;196:236--246.
\newblock doi:{10.1016/j.cpc.2015.06.010}.

\bibitem{Sherman2006}
Sherman W, Day T, Jacobson MP, Friesner RA, Farid R.
\newblock {Novel procedure for modeling ligand/receptor induced fit effects}.
\newblock J Med Chem. 2006;49(2):534--553.
\newblock doi:{10.1021/jm050540c}.

\bibitem{Jacobson:Pincus:Rapp:Day:Honig:Shaw:Friesner:2003}
Jacobson MP, Pincus DL, Rapp CS, Day TJF, Honig B, Shaw DE, et~al.
\newblock {A hierarchical approach to all-atom loop prediction}.
\newblock Proteins. 2004;55:351--367.

\bibitem{Jacobson2002}
Jacobson MP, Friesner RA, Xiang Z, Honig B.
\newblock {On the role of the crystal environment in determining protein
  side-chain conformations}.
\newblock J Mol Biol. 2002;320(3):597--608.
\newblock doi:{10.1016/S0022-2836(02)00470-9}.

\bibitem{AmberMD2013}
Salomon-Ferrer R, Case DA, Walker RC.
\newblock {An overview of the Amber biomolecular simulation package}.
\newblock Wiley Interdiscip Rev Comput Mol Sci. 2013;3(2):198--210.
\newblock doi:{10.1002/wcms.1121}.

\bibitem{Maier2015}
Maier JA, Martinez C, Kasavajhala K, Wickstrom L, Hauser KE, Simmerling C.
\newblock {ff14SB: Improving the Accuracy of Protein Side Chain and Backbone
  Parameters from ff99SB}.
\newblock J Chem Theory Comput. 2015;11(8):3696--3713.
\newblock doi:{10.1021/acs.jctc.5b00255}.

\bibitem{friesner2004glide}
Friesner RA, Banks JL, Murphy RB, Halgren TA, Klicic JJ, Mainz DT, et~al.
\newblock {Glide: A New Approach for Rapid, Accurate Docking and Scoring. 1.
  Method and assessmnet of Docking Accuracy}.
\newblock J Med Chem. 2004;47(7):1739--1749.

\bibitem{shelley2007epik}
Shelley JC, Cholleti A, Frye LL, Greenwood JR, Timlin MR, Uchimaya M.
\newblock {Epik: A software program for pKa prediction and protonation state
  generation for drug-like molecules}.
\newblock J Comput Aided Mol Des. 2007;21(12):681--691.
\newblock doi:{10.1007/s10822-007-9133-z}.

\bibitem{Gallicchio2012a}
Gallicchio E, Levy RM.
\newblock {Prediction of SAMPL3 host-guest affinities with the binding energy
  distribution analysis method (BEDAM)}.
\newblock J Comput Aided Mol Des. 2012;26(5):505--516.
\newblock doi:{10.1007/s10822-012-9552-3}.

\bibitem{GallicchioSAMPL4}
Gallicchio E, Deng N, He P, Wickstrom L, Perryman AL, Santiago DN, et~al.
\newblock {Virtual screening of integrase inhibitors by large scale binding
  free energy calculations: the SAMPL4 challenge.}
\newblock J Comput Aided Mol Des. 2014;28:475--490.
\newblock doi:{10.1007/s10822-014-9711-9}.

\bibitem{GPCRFilizola2011}
Provasi D, Artacho MC, Negri A, Mobarec JC, Filizola M.
\newblock {Ligand-Induced modulation of the Free-Energy landscape of G
  protein-coupled receptors explored by adaptive biasing techniques}.
\newblock PLoS Comput Biol. 2011;7(10):e1002193.
\newblock doi:{10.1371/journal.pcbi.1002193}.

\bibitem{Michino2017}
Michino M, Boateng CA, Donthamsetti P, Yano H, Bakare OM, Bonifazi A, et~al.
\newblock {Toward understanding the structural basis of partial agonism at the
  dopamine D3 receptor}.
\newblock J Med Chem. 2017;60(2):580--593.
\newblock doi:{10.1021/acs.jmedchem.6b01148}.

\bibitem{Deng2009}
Deng Y, Roux B.
\newblock {Computations of standard binding free energies with molecular
  dynamics simulations}.
\newblock J Phys Chem B. 2009;113(8):2234--2246.
\newblock doi:{10.1021/jp807701h}.

\bibitem{BruceMacdonald2018}
{Bruce Macdonald} HE, Cave-Ayland C, Ross GA, Essex JW.
\newblock {Ligand Binding Free Energies with Adaptive Water Networks:
  Two-Dimensional Grand Canonical Alchemical Perturbations}.
\newblock J Chem Theory Comput. 2018;14(12):6586--6597.
\newblock doi:{10.1021/acs.jctc.8b00614}.

\bibitem{Clark2009a}
Clark M, Meshkat S, Wiseman JS.
\newblock {Grand Canonical Free-Energy Calculations of Protein-Ligand Binding}.
\newblock J Chem Inf Model. 2009;49(4):934--943.
\newblock doi:{10.1021/ci8004397}.

\end{thebibliography}

\section*{Supporting Information}

\textbf{S1 Text. Chemistry and Synthesis}\\
\textbf{SF1 Fig. Synthesis of C3 analogues of (-)-stepholidine}\\
\textbf{ST1 Table. Experimental and calculated energetic quantities
with uncertainties}
\end{document}


\begin{center}
\textbf{\LARGE{}Supplementary Information:}
\par\end{center}{\LARGE \par}

\begin{center}
\textbf{\LARGE{}Inclusion of Enclosed Hydration Effects in the Binding
Free Energy Estimation of Dopamine D3 Receptor Complexes}
\par\end{center}{\LARGE \par}

\begin{center}
Rajat Kumar Pal$^{1,5}$, Steve Ramsey$^{2,5}$, Satishkumar Gadhiya$^{4,6}$,
Pierpaolo Cordone$^{4,5}$, \\
Lauren Wickstrom$^{3}$, Wayne W. Harding$^{4,5,6}$, Tom Kurtzman$^{2,5,6}$,
Emilio Gallicchio$^{1,5,6}$%
\footnote{Corresponding author: egallicchio@brooklyn.cuny.edu. $^{1}$Department
of Chemistry, Brooklyn College, 2900 Bedford Avenue, Brooklyn, New
York 11210. $^{2}$Department of Chemistry, Lehman College, The City
University of New York, 250 Bedford Park Blvd. West, Bronx, New York,
NY, 10468. $^{3}$Borough of Manhattan Community College, Department
of Science, The City University of New York, 199 Chambers Street,
New York, NY, 10007. $^{4}$Department of Chemistry, Hunter College,
The City University of New York, NY. $^{5}$Ph.D. Program in Biochemistry,
The Graduate Center of the City University of New York, New York,
NY 10016. $^{6}$Ph.D. Program in Chemistry, The Graduate Center of
the City University of New York, New York, NY 10016.%
}
\par\end{center}

\subsection*{S1 Chemistry and Synthesis}

\subsubsection*{4-(benzyloxy)-3-hydroxybenzaldehyde (5) }

To a stirring solution of 3,4-dihydroxybenzaldehyde, 4 (2.5 g, 18.1
mmol) in anhydrous Acetonitrile (50 mL), was added ${\rm K_{2}CO_{3}}$
(2.5 g, 18.1 mmol) followed by benzyl bromide (3.2 mL, 27.15 mmol)
slowly, at room temperature, under an inert atmosphere. Resulting
reaction mass was heated to reflux temperature and stirring was continued
for 2 h. The reaction solvent was removed by evaporation under reduced
pressure and to the resulted residue was added cold 10\% NaOH solution
and stirred for 10 min. and added the ethyl acetate (2 x 50 mL). Resulted
biphasic mixture was separated and aqueous layer was acidified using
4N HCl and extracted with DCM (3 x 30 mL), combined organic layer
was washed with brine solution, water, dried on ${\rm Na_{2}SO_{4}}$,
and concentrated under reduced pressure to obtain the residue, which
was purified by crystallization using ethylacetate to afford, 5 (2.7
g, 65\%); $^{1}{\rm H}$ NMR (400 MHz, ${\rm CDCl_{3}}$) $\delta$
9.83 (s, 1H ), 7.46-7.38 (m, 7H), 7.03 (d, $J$ = 8.2 Hz, 1H), 5.81
(s, 1H), 5.20 (s, 2H); $^{13}{\rm C}$ NMR (400 MHz, ${\rm CDCl_{3}}$)
$\delta$ 190.9, 150.9, 146.3, 135.2, 130.8, 128.9 X 2, 128.8, 127.9
X 2, 124.3, 114.4, 111.5, 71.2; HRMS (ESI) m/z calcd for ${\rm C_{14}H_{12}O_{3}}$
(${\rm [M+H]^{+}}$), 229.0859, found 229.0863.

\subsubsection*{2-(benzyloxy)-5-(2-nitrovinyl)phenoxy)(tert-butyl)diphenylsilane
(6)}

Compound 5 (2.7 g, 11.8 mmol) was dissolved in anhydrous DCM (50 mL).
Imidazole (1.6 g, 23.6 mmol) and DMAP (0.1 g, 0.12 mmol) were added
in to reaction mixture and allowed it to stir at 0\r{ }C for 10 minute.
To the reaction mixture, TBDPSCl (4.6 mL 17.7 mmol) was added and
the reaction was allowed to stir overnight at rt. Reaction mixture
was acidified with 2N HCl and extracted with (2 x 40 mL) DCM. Combined
organic layer was washed with brine solution, water, dried on ${\rm Na_{2}SO_{4}}$
and concentrated under reduced pressure to obtain the crude residue,
which was used in the next step without further purification. Residues
were dissolved in 50 mL acetic acid, ${\rm NH_{4}OAc}$ (1.0 , 12.98
mmol) and ${\rm CH_{3}NO_{2}}$ (2.5, 47.2 mmol) were added in it
and allowed it to reflux for 8 hr. Solvent was evaporated and the
residues were basified using 10\% NaOH and extracted with DCM (2 x
50 mL). Combined organic layer was dried over ${\rm Na_{2}SO_{4}}$,
and concentrated under reduced pressure to obtain a crude residue,
which was purified by column chromatography on silica gel using 10:90
EtOAC: Hexanes as eluent to afford compound 6 (5.1 g, 85\% over two
steps); ${\rm ^{1}H}$ NMR (400 MHz, ${\rm CDCl_{3}}$) $\delta$
7.73-7.63 (m, 5H), 7.46-7.27 (m, 11H), 7.05-7.00 (m, 2H), 6.86-6.77
(m, 2H), 4.97 (s, 2H), 1.10 (s, 9H); 13C NMR (400 MHz, ${\rm CDCl_{3}}$)
$\delta$ 153.5, 145.8, 139.1, 136.1 X 2, 135.4 X 2 , 135.3 X 2, 135.0,
134.8, 132.8, 130.1, 129.7, 128.5 X 3, 128.0 X 2, 127.8 X 4, 127.4
X 2, 124.8, 122.7, 119.5, 113.5, 70.5, 26.6; HRMS (ESI) m/z calcd
for ${\rm C_{31}H_{31}NO_{4}Si}$ ({[}M+H{]}$^{+}$), 510.2101, found
510.2098.

\subsubsection*{2-(4-(benzyloxy)-3-((tert-butyldiphenylsilyl)oxy)phenyl)ethan-1-amine
(7)}

In a dry round bottom flask, ${\rm LiBH_{4}}$ (1.8 g, 82.4 mmol),
was suspended in THF (30 mL) and TMSCl (5.3 mL, 41.2 mmol) was added
slowly. Solution of compound 6 (5.1 g, 10.3 mmol) in THF (30 mL) was
added slowly in to reaction mixture and allowed to reflux overnight.
Reaction was allowed to cool at rt and excess ${\rm LiBH_{4}}$ was
quenched with MeOH. 10\% NaOH (20 mL) was added to the reaction and
it was filtered through celite. Solvent was evaporated and the crude
mixture was extracted with DCM (2 x 50 mL). Combined organic layer
was dried over ${\rm Na_{2}SO_{4}}$, and concentrated under reduced
pressure to obtain compound 7; ${\rm ^{1}H}$ NMR (400 MHz, ${\rm CDCl_{3}}$)
$\delta$ 7.72-7.68 (m, 5H), 7.41-7.27 (m, 10H), 6.74 (d, J = 6.6
Hz, 1H) 6.62 (dd, J = 1.4, 6.5 Hz, 1H), 6.4 (d, J = 1.5 Hz, 1H), 4.88
(s, 2H), 2.72 (t, J = 5.8 Hz, 2H), 2.59 (t, J = 5.8 Hz, 2H), 1.08
(s, 9H); 13C NMR (400 MHz, ${\rm CDCl_{3}}$) \textgreek{d} 147.5,
144.4, 136.1, 134.4 X 2 , 135.3 X 2, 133.7, 132.2, 129.1, 128.7 X
3, 128.6, 127.2 X 4, 126.8 X 2, 126.7, 126.4 X 2, 120.7, 119.7, 113.2
X 2, 69.6, 40.9, 34.5, 25.5; HRMS (ESI) m/z calcd for ${\rm C_{31}H_{35}NO_{2}Si}$
({[}M+H{]}$^{+}$), 482.2515, found 482.2509.

\subsubsection*{2-(4-(benzyloxy)-2-(hydroxymethyl)-3-methoxyphenyl)-N-(4-(benzyloxy)-3-\protect \linebreak{}
((tert-butyldiphenylsilyl)oxy)phenethyl) acetamide (9)}

Compound 8 (1.5 g, 5.3 mmol) and primary amine 7 (2.5g, 5.3 mmol)
were refluxed in to ethanol (20 mL) for 16 h. Solvent was evaporated
and the residues were purified by column chromatography (2 \% MeOH/DCM)
to yield 10 (3.2g, 78\%). ${\rm ^{1}H}$ NMR (${\rm CDCl_{3}}$, 600
MHz) $\delta$ 7.69 (dd, J = 8.1, 1.3 Hz, 4H), 7.43-7.36 (m, 6H),
7.33-7.28 (m, 10H), 6.82 (d, J = 8.4 Hz, 1H), 6.72 (d, J = 8.4 Hz,
1H), 6.70 (d, J = 8.1 Hz, 1H), 6.42-6.38 (m, 2H), 5.56 (s, 1H), 5.10
(s, 2H), 4.93 (s, 2H), 4.64 (d, J = 6.2 Hz, 2H), 3.93 (s, 3H), 3.55
(t, J = 6.2 Hz, 1H), 3.35 (s, 2H), 3.11 (q, J = 6.4 Hz, 2H), 2.41
(t, J = 6.6 Hz, 2H), 1.09 (s, 9H); 13C NMR (${\rm CDCl_{3}}$, 150
MHz) $\delta$ 171.4, 151.1, 148.7, 145.5, 137.3, 137.0, 135.5 X 3,
134.2, 133.4, 131.4, 129.8 X 3, 128.6 X 3, 128.3 X 3, 128.0 X 3, 127.7
X 2, 127.6 X 2, 127.4 X 2, 127.2 X 2, 125.6, 121.8, 120.6, 114.3,
114.1, 70.8 X 2, 61.8, 56.6, 40.7, 40.6, 34.3, 26.6 X 3, 19.7; (ESI)
m/z calcd. for ${\rm C_{48}H_{51}NO_{6}Si}$ ({[}M+Na{]}$^{+}$),
788.3378, found 788.3416.

\subsubsection*{3-(benzyloxy)-6-(2-((4-(benzyloxy)-3-((tert-butyldiphenylsilyl)oxy)phenethyl)amino)-\protect \linebreak{}
2-oxoethyl)-2-methoxybenzyl acetate (10)}

To the solution of compound 10 (3.1 g, 4.1 mmol) in DCM (20 mL) was
added pyridine (0.6 mL, 8.2 mmol) and DMAP (10 mg, 0.08 mmol). The
mixture was cooled to 0\r{ } C and acetic anhydride (0.8 mL, 8.2 mmol)
was added dropwise. The solution was allowed to stir for 1.5 h at
room temperature, the reaction mixture was diluted with DCM (20 mL),
washed with 1 N HCl and dried over sodium sulfate and concentrated.
The residue was purified by flash chromatography (1.5\% MeOH/DCM)
to afford 10 (3.1 g, 92\%). ${\rm {\rm ^{1}H}}$ NMR (${\rm CDCl_{3}}$,
500 MHz) $\delta$ 7.66 (dd, J = 8.1, 1.3 Hz, 4H), 7.41-7.36 (m, 6H),
7.33-7.27 (m, 10H), 6.88 (d, J = 8.4 Hz, 1H), 6.78 (d, J = 8.4 Hz,
1H), 6.68 (d, J = 8.4 Hz, 1H), 6.40-6.38 (m, 2H), 5.14 (s, 2H), 5.08
(s, 2H), 4.88 (s, 2H), 3.88 (s, 3H), 3.45 (s, 2H), 3.10 (q, J = 6.7
Hz, 2H), 2.40 (t, J = 6.8 Hz, 2H), 1.97 (s, 3H), 1.07 (s, 9H);); ${\rm ^{13}C}$
NMR (${\rm CDCl_{3}}$, 125 MHz) $\delta$ 170.8, 170.5, 151.1, 149.2,
148.3, 145.4, 137.2, 136.6, 135.4 X 3, 133.3, 131.2, 129.7 X 3, 128.6
X 3, 128.3 X 3, 128.0 X 3, 127.6 X 4, 127.4 X 2, 127.2 X 2, 126.2,
121.6, 120.5, 114.8, 114.1, 70.7 X 2, 61.4, 58.4, 40.5, 40.4, 34.5,
26.6 X 3, 20.9, 19.7; (ESI) m/z calcd. for ${\rm C_{50}H_{53}NO_{7}Si}$
({[}M+Na{]}$^{+}$), 830.3489, found 830.3491.

\subsubsection*{(S)-3-(benzyloxy)-6-((7-(benzyloxy)-6-hydroxy-1,2,3,4-tetrahydroisoquinolin-1-yl)methyl)-\protect \linebreak{}
 2-methoxybenzyl acetate (11)}

To the solution of 10 (3.0 g, 3.7mmol) in dry acetonitrile (20 mL)
was added ${\rm POCl_{3}}$ (2.1 mL, 22.3 mmol) and the mixture was
refluxed for 1 h. The reaction mixture was cooled to room temperature
and the solvent was evaporated. The residue was dissolved in DCM (30
mL) and washed with saturated ${\rm NaHCO_{3}}$ (15 mL). The organic
layer was dried over sodium sulfate and evaporated to dryness. The
yellow solid residue was dissolved in DMF (15 mL) and RuCl{[}(R,R)-TsDPEN(P-cymene){]}
(20 mg, 0.03 mmol) was added under nitrogen. The solution was purged
Nitrogen gas for 15 minutes. The mixture of formic acid (0.8 mL, 20.7
mmol ) and triethyl amine (0.3 mL 2.21 mmol) was added in to the reaction
mixture and allowed it to stir overnight at rt. The reaction mixture
was adjusted pH to 8 with saturated ${\rm NaHCO_{3}}$ and extracted
with ethyl acetate (3 X 35 mL). The combined organic layer was washed
with brine, dried over ${\rm Na_{2}SO_{4}}$ and concentrated. The
crude product was purified by column chromatography on silica gel
using 3:97 to 5:95 MeOH:DCM as eluent to afford compound, 11 (1.4
g, 67\% from two steps).

\subsubsection*{(S)-2,10-bis(benzyloxy)-9-methoxy-5,8,13,13a-tetrahydro-6H-isoquinolino{[}3,2-a{]}isoquinolin-3-ol
(12)}

Compound 11 (1.4 g, 2.5 mmol) was dissolved in ethanol (20 mL) and
added 10\% NaOH (10 mL) to the solution then the mixture was refluxed
for 1 h. The mixture was cooled and ethanol was evaporated. The residue
was extracted with ehyl acetate (2 X 30 mL) and the combined organic
layer was washed dried over ${\rm Na_{2}SO_{4}}$ and evaporated to
dryness. The residues were dissolved in dry DCM (15 mL) and the solution
was cooled 0\r{ } C. Thionyl chloride (0.9 mL, 12.5 mmol) was added
dropwise to the reaction mixture and allowed it to stir for 1 hr.
Reaction solution was cooled to 0\r{ } C and saturated ${\rm NaHCO_{3}}$
(30 mL) was added to it and allowed it stir for 1 h. Reaction mixture
was extracted with DCM (2 X 50 mL), combined organic layer was dried
over sodium sulfate and evaporated to dryness. The crude product was
purified by column chromatography on silica gel using 1:99 MeOH:DCM
as eluent to afford compound, 12 (1.1 g, 88\% over two steps).${\rm ^{1}H}$
NMR (${\rm CDCl_{3}}$, 500 MHz) $\delta$ 7.45-7.37 (m, 10H), 6.83
(s, 2H), 6.78 (s, 1H), 6.71 (s, 1H), 5.58 (s, 1H), 5.14-5.06 (m, 4H),
4.25 (d, J = 15.8 Hz, 1H), 3.90 (s, 3H), 3.55-3.52 (m, 2H), 3.20-3.08
(m, 3H), 2.80 (dd, J = 11.5, 15.4 Hz, 1H), 2.67-2.60 (m, 2H); ${\rm ^{13}C}$
NMR (${\rm CDCl_{3}}$, 120 MHz) $\delta$ 149.3, 145.6, 144.3, 144.2,
137.2, 136.4, 129.2, 128.9, 128.7 X 2, 128.5 X 2, 128.4, 128.2, 128.0,
127.9, 127.8 X 2, 127.3 X 2, 123.7, 114.4, 113.1, 109.4, 71.4, 70.9,
60.3, 59.3, 54.0, 51.5, 36.5, 28.9; (ESI) m/z calcd. for ${\rm C_{32}H_{31}NO_{4}}$
({[}M+H{]}$^{+}$), 494.2326, found 494.2330.

\subsubsection*{General synthetic procedure for the compounds 1a-1f as demonstrated
for 1a (S)-3-ethoxy-9-methoxy-5,8,13,13a-tetrahydro-6H-isoquinolino{[}3,2-a{]}isoquinoline-2,10-diol
(1a)}

Compound, 12 (1.0 eq) was dissolved in anhydrous DMF and added ${\rm K_{2}CO_{3}}$
(2.0 eq) followed by ethyl bromide (1.2 eq). Resulted reaction mass
was allowed to stir at rt for 5 h. The reaction mass was quenched
with cold water and extracted with ethyl acetate (3 x 50 mL). Combined
organic layer was washed with brine solution, dried over ${\rm Na_{2}SO_{4}}$
and concentrated to obtain a residue, which was used in the next step
without further purification. Residues were dissolved in ethanol (10
mL) and refluxed with conc. HCl (4 mL) for 3 hr. The reaction mixture
was concentrated in vacuo and residue was basified using aqueous ammonia
solution and extracted with ethyl acetate (2 X 20 mL). Combined organic
layer was washed with brine solution, dried over ${\rm Na_{2}SO_{4}}$
and concentrated under reduced pressure to obtain a residue, which
was purified by column chromatography on silica gel using 2:98 MeOH:DCM
as eluent to afford compound 1a. Yield 71 \%, white solid, Mp 104
\textsuperscript{0}C; $^{1}{\rm H}$ NMR (${\rm CDCl_{3}}$, 500 MHz)
$\delta$ 6.84-6.79 (m, 3H), 6.59 (s, 1H), 4.20 (d, J = 15.4 Hz, 1H),
4.10 (q, J = 7.0 Hz, 2H), 3.81 (s, 3H), 3.58-3.53 (m, 2H), 3.27-3.09
(m, 3H), 2.80 (dd, J = 11.4, 15.9 Hz, 1H), 2.67-2.62 (m, 2H), 1.45
(t, J = 7.0, 3H); $^{13}{\rm C}$ NMR (${\rm CDCl_{3}}$, 125 MHz)
$\delta$ 146.4, 144.3, 144.0, 143.1, 130.3, 127.9, 127.3, 125.8,
125.0, 114.0, 111.4, 111.2, 64.4, 60.7, 59.3, 53.9, 51.7, 36.2, 29.1,
14.0; (ESI) m/z calcd. for ${\rm C_{20}H_{23}NO_{4}}$ ({[}M+H{]}$^{+}$),
342.1700, found 342.1697.

\subsubsection*{(S)-9-methoxy-3-propoxy-5,8,13,13a-tetrahydro-6H-isoquinolino{[}3,2-a{]}
isoquinoline-2,10-diol (1b) }

Yield 75\%, dark brown solid, Mp 82 \textsuperscript{0}C; $^{1}{\rm H}$
NMR (${\rm CDCl_{3}}$, 500 MHz) $\delta$ 6.85-6.81 (m, 3H), 6.59
(s, 1H), 4.28 (d, J = 15.4 Hz, 1H), 3.99 (t, J = 6.6 Hz, 2H), 3.82
(s, 3H), 3.65-3.58 (m, 2H), 3.28-3.24 (m, 2H), 3.15 (td, J = 5.6,
16.7 Hz, 1H), 2.84 (dd, J = 11.8, 16.0 Hz, 1H), 2.70-2.66 (m, 2H),
1.84 (sext, J = 7.1 Hz, 2H), 1.05 (t, J = 7.5, 3H); $^{13}{\rm C}$
NMR (${\rm CDCl_{3}}$, 125 MHz) $\delta$ 146.5, 144.5, 144.1, 143.2,
130.3, 127.9, 127.3, 125.8, 125.0, 114.2, 111.5, 111.2, 70.4, 60.7,
59.4, 53.9, 51.7, 36.1, 29.1, 22.6, 10.5; (ESI) m/z calcd. for ${\rm C_{21}H_{25}NO_{4}}$
({[}M+H{]}$^{+}$), 356.1856, found 356.1849.

\subsubsection*{(S)-3-butoxy-9-methoxy-5,8,13,13a-tetrahydro-6H-isoquinolino{[}3,2-a{]}
isoquinoline-2,10-diol (1c)}

Yield 76\%, yellowish powder, Mp 85 \textsuperscript{0}C; $^{1}{\rm H}$
NMR (${\rm CDCl_{3}}$, 500 MHz) $\delta$ 6.85-6.81 (m, 3H), 6.59
(s, 1H), 4.28 (d, J = 15.4 Hz, 1H), 3.99 (t, J = 6.6 Hz, 2H), 3.82
(s, 3H), 3.65-3.58 (m, 2H), 3.28-3.24 (m, 2H), 3.15 (td, J = 5.6,
16.7 Hz, 1H), 2.84 (dd, J = 11.8, 16.0 Hz, 1H), 2.70-2.66 (m, 2H),
1.84 (sext, J = 7.1 Hz, 2H), 1.05 (t, J = 7.5, 3H); $^{13}{\rm C}$
NMR (CDCl3, 125 MHz) $\delta$ 146.5, 144.5, 144.0, 143.1, 130.2,
127.8, 127.3, 125.8, 125.0, 114.2, 111.4, 111.1, 68.6, 60.7, 59.4,
53.9, 51.7, 36.1, 31.3, 29.1, 19.2, 13.8; (ESI) m/z calcd. for ${\rm C_{22}H_{27}O_{4}}$
({[}M+H{]}$^{+}$), 370.2013, found 370.2006.

\subsubsection*{(S)-9-methoxy-3-(pentyloxy)-5,8,13,13a-tetrahydro-6H-isoquinolino{[}3,2-a{]}
isoquinoline-2,10-diol (1d)}

Yield 73\%, brown solid, Mp 73 \textsuperscript{0}C; $^{1}{\rm H}$
NMR (${\rm CDCl_{3}}$, 500 MHz) $\delta$ 6.83-6.79 (m, 3H), 6.59
(s, 1H), 5.56 (s, 1H), 4.20(d, J = 15.3 Hz, 1H), 4.02 (t, J = 6.6
Hz, 2H), 3.81 (s, 3H), 3.58-3.55 (m, 2H), 3.27-3.10 (m, 3H), 2.80
(dd, J = 11.6, 15.8 Hz, 1H), 2.67-2.62 (m, 2H), 1.82 (pent, J = 6.6
Hz, 2H), 1.48-1.36 (m, 4H), 0.94 (t, J = 7.2, 3H); $^{13}{\rm C}$
NMR (${\rm CDCl_{3}}$, 125 MHz) $\delta$ 146.4, 144.5, 144.1, 143.1,
130.2, 127.8, 127.3, 125.8, 125.0, 114.1, 111.4, 111.1, 68.9, 60.7,
59.4, 53.9, 51.7, 36.1, 29.0, 28.9, 28.2, 22.4, 14.0 ; (ESI) m/z calcd.
for ${\rm C_{23}H_{29}NO_{4}}$ ({[}M+H{]}$^{+}$), 384.2169, found
384.2165.

\subsubsection*{(S)-3-(hexyloxy)-9-methoxy-5,8,13,13a-tetrahydro-6H-isoquinolino{[}3,2-a{]}
isoquinoline-2,10-diol (1e)}

Yield 70\%, dark brown powder, Mp 61 \textsuperscript{0}C; $^{1}{\rm H}$
NMR (${\rm CDCl_{3}}$, 500 MHz) $\delta$ 6.83-6.79 (m, 3H), 6.59
(s, 1H), 4.22 (d, J = 15.3 Hz, 1H), 4.02 (t, J = 6.6 Hz, 2H), 3.81
(s, 3H), 3.58(d, J = 14.7 Hz, 2H), 3.28-3.12 (m, 3H), 2.84-2.79 (m,
1H), 2.69-2.64 (m, 2H), 1.81 (pent, J = 6.7 Hz, 2H), 1.49-1.43 (m,
2H), 1.34 (sext, J = 3.7, 4H), 0.91 (t, J = 7.0, 3H); $^{13}{\rm C}$
NMR (${\rm CDCl_{3}}$, 125 MHz) $\delta$ 146.4, 144.5, 144.0, 143.1,
130.2, 127.9, 127.3, 125.8, 125.0, 114.1, 111.4, 111.1, 68.9, 60.7,
59.4, 53.9, 51.7, 36.2, 31.5, 29.2, 29.1, 25.7, 22.6, 14.0; (ESI)
m/z calcd. for ${\rm C_{24}H_{31}NO_{4}}$ ({[}M+Na{]}$^{+}$), 398.2326,
found 398.2319.

\subsubsection*{(S)-3-(2-fluoroethoxy)-9-methoxy-5,8,13,13a-tetrahydro-6H-isoquinolino{[}3,2-a{]}
isoquinoline-2,10-diol (1f)}

Yield 87\%, yellow powder, Mp 95 \textsuperscript{0}C; $^{1}{\rm H}$
NMR (${\rm CDCl_{3}}$, 500 MHz) $\delta$ 6.85-6.79 (m, 3H), 6.61
(s, 1H), 4.82 (t, J = 3.9 Hz, 1H), 4.73 (t, J = 3.9 Hz, 1H), 4.32-4.19
(m, 3H), 3.81 (s, 3H), 3.58-3.54 (m, 2H), 3.27-3.10 (m, 3H), 2.80
(dd, J = 11.5, 15.7 Hz, 1H), 2.68-2.62 (m, 2H); $^{13}{\rm C}$ NMR
(${\rm CDCl_{3}}$, 125 MHz) $\delta$ 146.4, 144.3, 143.8, 143.1,
131.5, 127.8, 127.2, 126.0, 125.0, 114.2, 112.4, 111.9, 82.4, 81.0,
68.5, 68.3, 60.7, 59.3, 53.8, 51.6, 36.0, 29.0; (ESI) m/z calcd. for
C20H22FNO4 ${\rm C_{20}H_{22}FNO_{4}}$ ({[}M+Na{]}$^{+}$), 360.1606,
found 360.1599.

\subsubsection*{(\textpm{})-Stepholidine (2.1)}

Lithium aluminum hydride (40 mg, 1.1 mmol) was stirred in THF at 0\r{ }
C for 10 min. Compound 3.11 (0.3 g, 0.7 mmol) in THF was added to
the reaction mixture at 0\r{ } C and it was allowed to reflux for
2 h. The reaction mixture was allowed to cool to rt and excess of
lithium aluminum hydride was quenched with water. THF was evaporated
and crude mixture was extracted with dichloromethane. The organic
layer was dried over sodium sulfate, filtered and evaporated to dryness.
The crude product was refluxed in mixture of methanol (5 mL) and concentrated
hydrochloric acid (5 mL) for 3 h. The reaction mixture was basified
by ammonia solution and extracted with dichloromethane (20 mL \texttimes{}
2). The combined organic layer was dried over sodium sulfate and concentrated
to give the crude product, which was purified by flash column chromatography
on silica gel (2\%-6\% methanol/dichloromethane) to give compound
2.1 (0.29 g, 81\% over two steps): Yellowish white crystals, mp 157\r{ }
C; $^{1}{\rm H}$ NMR (${\rm CDCl_{3}}$, 400MHz) $\delta$ 6.83-6.78
(m, 3H), 6.60 (s, 1H), 4.20 (d, J = 15.8 Hz, 1H), 3.87 (s, 3H), 3.81
(s, 3H), 3.58-3.53 (m, 2H), 3.27-3.10 (m, 3H), 2.80 (dd, J = 15.9,
11.4 Hz, 1H), 2.70-2.62 (m, 2H); $^{13}{\rm C}$ NMR (${\rm CDCl_{3}}$,
400MHz) $\delta$ 146.4, 145.1, 144.0, 143.1, 130.4, 127.8, 127.3,
125.9, 125.0, 114.1, 111.3, 110.6, 60.7, 59.3, 55.9, 58.9, 51.7, 36.1,
29.1; HRMS (ESI) m/z calcd. for C19H21NO4 ${\rm C_{19}H_{21}NO_{4}}$
({[}M+H{]}$^{+}$), 328.1551, found 328.1543.

\setcounter{table}{0}\renewcommand{\thetable}{ST\arabic{table}}

\setcounter{figure}{0}\renewcommand{\thefigure}{SF\arabic{figure}}

\begin{figure}
\centering{}\includegraphics[scale=0.7]{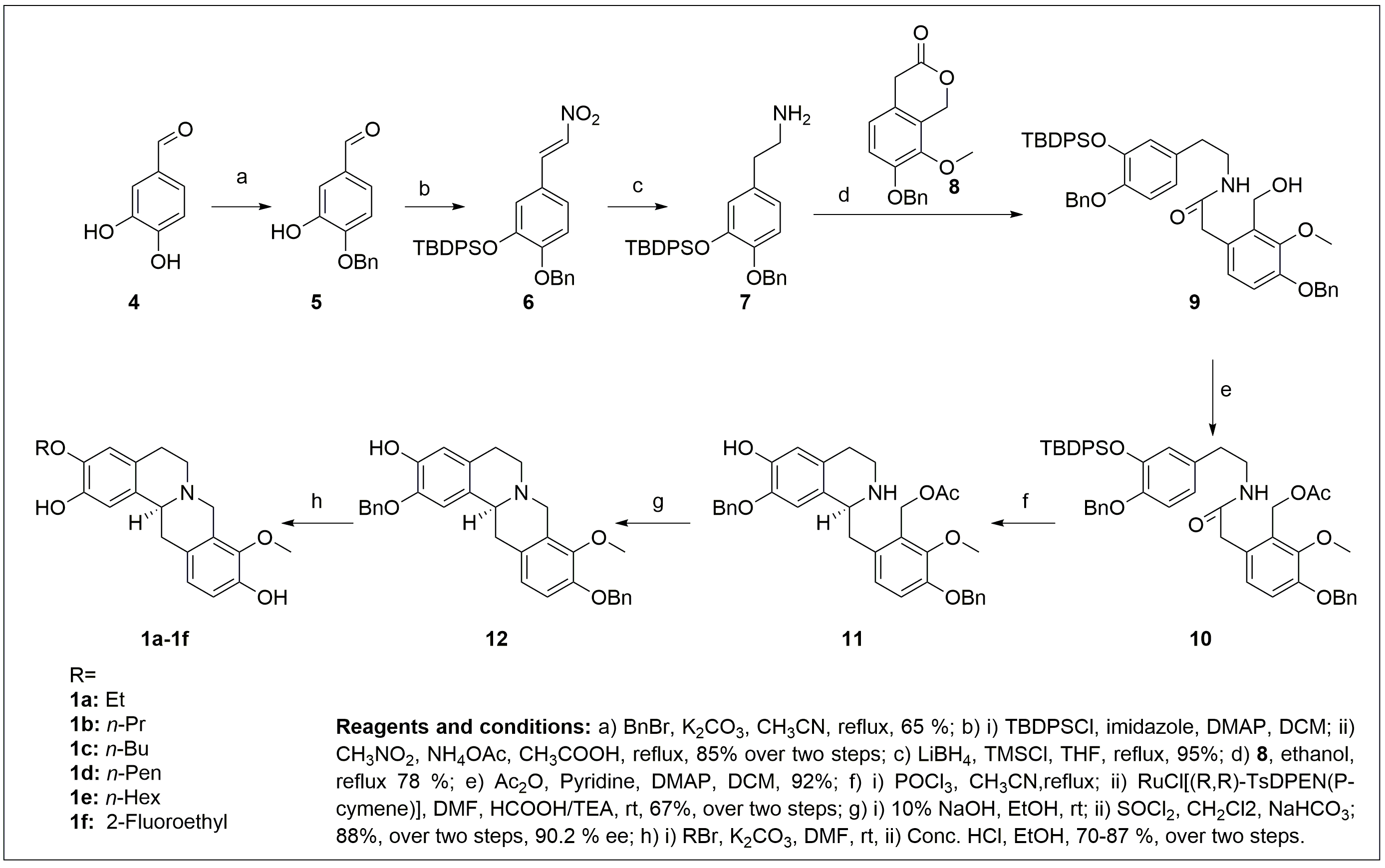}\caption{Synthesis of C3 analogues of (-)-stepholidine}
\end{figure}

\begin{sidewaystable}
\caption{Experimental and calculated energetic quantities with uncertainties.
Experimental and calculated binding free energies, average binding
energies and reorganization free energies of the (-)-stepholidine
C3 analogues with and without enclosed hydration corrections.}
\medskip{}

\raggedright{}{\footnotesize{}}%
\begin{tabular*}{1\columnwidth}{@{\extracolsep{\fill}}c>{\centering}m{0.12\textwidth}ccccccc}
\toprule 
{\footnotesize{}Ligands} & {\footnotesize{}Structure} & {\footnotesize{}$\Delta G_{{\rm exp}}^{\circ^{a}}$} & \multicolumn{3}{c}{{\footnotesize{}Without enclosed hydration model}} & \multicolumn{3}{c}{{\footnotesize{}With enclosed hydration model}}\tabularnewline
\cmidrule{4-9} 
 &  &  & {\footnotesize{}$\Delta G_{{\rm calc}}^{\circ^{a}}$} & {\footnotesize{}$\Delta E_{{\rm b}}^{^{a}}$} & {\footnotesize{}$\Delta G_{{\rm reorg}}^{\circ^{a}}$} & {\footnotesize{}$\Delta G_{{\rm calc}}^{\circ^{a}}$} & {\footnotesize{}$\Delta E_{{\rm b}}^{^{a}}$} & {\footnotesize{}$\Delta G_{{\rm reorg}}^{\circ^{a}}$}\tabularnewline
\midrule
{\footnotesize{}1a} & %
\begin{minipage}[c]{0.1\textwidth}%
\includegraphics[scale=0.25]{Extra_figures/table4_column2_row3}%
\end{minipage} & {\footnotesize{}$-10.1$} & {\footnotesize{}$-2.2\pm0.07$} & {\footnotesize{}$-36.9\pm0.22$} & {\footnotesize{}$34.7\pm0.24$} & {\footnotesize{}$-8.8\pm0.07$} & {\footnotesize{}$-42.5\pm0.21$} & {\footnotesize{}$33.7\pm0.23$}\tabularnewline
{\footnotesize{}1b} & %
\makebox[0.1\textwidth][l]{%
\includegraphics[scale=0.25]{Extra_figures/table4_column2_row4}%
} & {\footnotesize{}$-10.0$} & {\footnotesize{}$-2.3\pm0.07$} & {\footnotesize{}$-38.0\pm0.22$} & {\footnotesize{}$35.7\pm0.23$} & {\footnotesize{}$-10.4\pm0.07$} & {\footnotesize{}$-44.7\pm0.19$} & {\footnotesize{}$34.3\pm0.20$}\tabularnewline
{\footnotesize{}1c} & %
\begin{minipage}[t]{0.1\textwidth}%
\includegraphics[scale=0.25]{Extra_figures/table4_column2_row5}%
\end{minipage} & {\footnotesize{}$-10.0$} & {\footnotesize{}$-1.8\pm0.08$} & {\footnotesize{}$-40.3\pm0.23$} & {\footnotesize{}$38.5\pm0.24$} & {\footnotesize{}$-11.5\pm0.08$} & {\footnotesize{}$-48.1\pm0.18$} & {\footnotesize{}$36.6\pm0.19$}\tabularnewline
{\footnotesize{}1d} & %
\begin{minipage}[t]{0.12\textwidth}%
\includegraphics[scale=0.25]{Extra_figures/table4_column2_row6}%
\end{minipage} & {\footnotesize{}$-10.2$} & {\footnotesize{}$-0.3\pm0.09$} & {\footnotesize{}$-43.7\pm0.26$} & {\footnotesize{}$43.4\pm0.29$} & {\footnotesize{}$-10.6\pm0.08$} & {\footnotesize{}$-55.6\pm0.22$} & {\footnotesize{}$45.0\pm0.24$}\tabularnewline
{\footnotesize{}1e} & %
\begin{minipage}[t]{0.12\textwidth}%
\includegraphics[scale=0.25]{Extra_figures/table4_column2_row7}%
\end{minipage} & {\footnotesize{}$-10.4$} & {\footnotesize{}$-3.9\pm0.1$1} & {\footnotesize{}$-39.6\pm0.21$} & {\footnotesize{}$35.7\pm0.23$} & {\footnotesize{}$-12.5\pm0.08$} & {\footnotesize{}$-55.2\pm0.22$} & {\footnotesize{}$42.7\pm0.23$}\tabularnewline
{\footnotesize{}1f} & %
\begin{minipage}[t]{0.12\textwidth}%
\includegraphics[scale=0.25]{Extra_figures/table4_column2_row8}%
\end{minipage} & {\footnotesize{}$-9.64$} & {\footnotesize{}$-3.1\pm0.1$1} & {\footnotesize{}$-32.7\pm0.27$} & {\footnotesize{}$29.6\pm0.29$} & {\footnotesize{}$-8.9\pm0.07$} & {\footnotesize{}$-43.2\pm0.19$} & {\footnotesize{}$34.3\pm0.20$}\tabularnewline
\midrule 
\multicolumn{9}{l}{%
\begin{minipage}[t]{1\columnwidth}%
\begin{flushleft}
{\scriptsize{}$^{a}$All values in kcal/mol. }
\par\end{flushleft}%
\end{minipage}}\tabularnewline
\bottomrule
\end{tabular*}
\end{sidewaystable}